\documentclass[11pt]{article}
\usepackage{amsmath,amsfonts,amssymb,graphicx,epsfig}
\usepackage[usenames]{color}
\usepackage{pstricks}
\numberwithin{equation}{section}

\newcommand{\nc}{\newcommand}
\nc{\fh}{\hat{f}}
\nc{\muh}{\hat{\mu}}
\nc{\nuh}{\hat{\nu}}
\nc{\bib}{\bibitem}
\nc{\al}{\alpha}
\nc{\g}{\gamma}
\nc{\G}{\Gamma}
\nc{\D}{\Delta}
\nc{\eps}{\epsilon}
\nc{\la}{\lambda}
\nc{\La}{\Lambda}
\nc{\var}{\varphi}
\nc{\pa}{\partial}
\nc{\nn}{\nonumber \\ }
\nc{\hf}{\frac{1}{2}}
\nc{\dz}{\frac{dz}{2\pi i}}
\nc{\bin}[2]{\left(\!\!\!\begin{array}{c} {#1}\\ {#2} \end{array}\!\!\!\right)}
\nc{\be}{\begin{equation}}
\nc{\ee}{\end{equation}}
\nc{\bea}{\begin{eqnarray}}
\nc{\eea}{\end{eqnarray}}
\nc{\bra}[1]{\langle {#1}|}
\nc{\ket}[1]{|{#1}\rangle}
\nc{\chit}{\raisebox{0.25ex}{$\chi$}}
\nc{\Db}{\mbox{\boldmath $D$}}
\nc{\Hb}{\mbox{\boldmath $H$}}
\nc{\Hc}{{\cal H}}
\nc{\Rc}{{\cal R}}
\nc{\Lc}{{\cal L}}
\nc{\Vc}{{\cal V}}
\nc{\Ib}{\mbox{\boldmath $I$}}
\nc{\qb}{\bar{q}}
\def\vvdots{\mathinner{\mkern1mu\raise1pt\vbox{\kern7pt\hbox{.}}\mkern2mu
  \raise4pt\hbox{.}\mkern2mu\raise7pt\hbox{.}\mkern1mu}}
\nc{\gauss}[2]{\left[\!\!\begin{array}{c} {#1}\\ {#2} \end{array}\!\!\right]}
\nc{\sbin}[2]{\left\{\!\!\!\begin{array}{c} {#1}\\ {#2} 
\end{array}\!\!\!\right\}}
\nc{\sbinlr}[2]{\Big\langle\!\!\begin{array}{c} {#1}\\ {#2} 
\end{array}\!\!\Big\rangle}
\nc{\bino}[2]{\left(\!\!\begin{array}{c} {#1}\\ {#2} \end{array}\!\!\right)}
\definecolor{lightblue}{rgb}{.61,.61,1}
\definecolor{midblue}{rgb}{.7,.7,1}
\definecolor{lightlightblue}{rgb}{.85,.85,1}
\definecolor{lightestblue}{rgb}{.96,.96,1}
\nc{\ch}{{\rm ch}}
\nc{\R}{{\cal R}}
\nc{\dkk}{\delta_{j,\{k,k'\}}^{(2)}}
\nc{\ddkk}{\delta_{j,\{k,k'\}}^{(4)}}
\nc{\dddkk}{\delta_{j,\{k,k'\}}^{(8)}}
\nc{\dnn}{\delta_{j,\{n,n'\}}^{(2)}}
\nc{\ddnn}{\delta_{j,\{n,n'\}}^{(4)}}
\nc{\dddnn}{\delta_{j,\{n,n'\}}^{(8)}}

\begin{document}

\topmargin -5mm
\oddsidemargin 5mm

\begin{titlepage}
\setcounter{page}{0}

\vspace{8mm}
\begin{center}
{\Large {\bf Fusion Algebras of Logarithmic Minimal Models}}

\vspace{10mm}
{\Large J{\o}rgen Rasmussen}\ \ and\ \ {\Large Paul A. Pearce}\\[.3cm]
{\em Department of Mathematics and Statistics, University of Melbourne}\\
{\em Parkville, Victoria 3010, Australia}\\[.4cm]
J.Rasmussen@ms.unimelb.edu.au,\quad P.Pearce@ms.unimelb.edu.au

\end{center}

\vspace{8mm}
\centerline{{\bf{Abstract}}}
\vskip.4cm
\noindent
We present explicit conjectures for the chiral fusion algebras of the logarithmic minimal models 
${\cal LM}(p,p')$ considering 
Virasoro representations with no enlarged or extended symmetry algebra. 
The generators of fusion are countably infinite in number but
the ensuing fusion rules are quasi-rational in the sense that the fusion of a finite number of 
representations decomposes into a finite direct sum of representations. 
The fusion rules are 
commutative, associative and exhibit an $s\ell(2)$ structure
but require so-called Kac representations which are reducible yet indecomposable representations
of rank 1. In particular, the identity of the fundamental fusion algebra
is in general a reducible yet indecomposable Kac representation of rank 1.
We make detailed comparisons of our fusion rules 
with the results of Gaberdiel and Kausch for $p=1$ and with Eberle and Flohr for 
$(p,p')=(2,5)$ corresponding to the logarithmic Yang-Lee model. 
In the latter case, we confirm the appearance of indecomposable representations of rank 3.  
We also find that closure of a fundamental fusion algebra is achieved
without the introduction of indecomposable representations of rank higher than 3.
The conjectured fusion rules are supported, within our lattice approach, by extensive numerical 
studies of  
the associated integrable lattice models. Details of our lattice findings and numerical results
will be presented elsewhere. The agreement of our fusion rules with the previous fusion rules lends 
considerable support for the identification of the logarithmic minimal models ${\cal LM}(p,p')$ 
with the augmented $c_{p,p'}$ (minimal) models defined algebraically.

\end{titlepage}
\newpage
\renewcommand{\thefootnote}{\arabic{footnote}}
\setcounter{footnote}{0}

\section{Introduction}

Since the seminal work of Gaberdiel and Kausch~\cite{GK96}, steady progress has been made in 
understanding the fusion algebras of logarithmic Conformal Field Theories (CFTs)~\cite{Flohr03,Gaberdiel03}.
Three key approaches to these problems are the algebraic, supergroup and lattice approaches.
Within the algebraic approach, the so-called augmented $c_{p,p'}$ (minimal) models, 
with $p,p'$ coprime integers, have been extensively studied. These logarithmic theories are characterized directly by algebraic properties of the CFT. 
Initially, work focused~\cite{GK96} on the case $p=1$ but recently Eberle and Flohr~\cite{EF06} 
extended the application of the Nahm algorithm~\cite{Nahm94} to obtain fusion rules level-by-level for some $p>1$. Nevertheless, it seems fair to say that knowledge in the general case remains limited. 
The supergroup approach originated with Rozansky and Saleur~\cite{RS92} 
but has since been pursued by other authors as well, 
see~\cite{QS07} and references therein. In this approach, reducible yet 
indecomposable representations arise somewhat
automatically as a consequence of the supergroup structure. 
The lattice approach underlying the present work was initiated in \cite{PRZ}. 
Within this approach, the logarithmic minimal models ${\cal LM}(p,p')$ 
are defined as CFTs via the continuum scaling limit of a series of Yang-Baxter integrable lattice models. 
The lattice approach has the advantage of being firmly rooted in physical origins but the disadvantage that the algebraic properties of the resulting CFTs are not readily accessible.

It would appear natural to identify the $c_{p,p'}$ and ${\cal LM}(p,p')$ models and to expect these theories to 
play the same role for logarithmic CFTs that the usual minimal models~\cite{BPZ84} do for rational CFTs. Indeed, based on comparison of conformal data and fusion rules, it seems that the first members of these series with $(p,p')=(1,2)$, $(2,3)$ corresponding to critical dense polymers~\cite{PR06} and critical percolation~\cite{RP07}, respectively, 
do in fact coincide. In general, however, care needs to be exercised. For rational CFTs, there is a precise axiomatic definition~\cite{MooreSeiberg} of a rational CFT. 
In practice, this means that these theories can be classified by a certain set of data such as the central charge, conformal weights, characters of irreducible representations, modular invariant partition functions on the torus and operator-product expansions. In contrast, as of now,
no such general theory exists for logarithmic CFTs. 
To the contrary, at least in principle, it is possible for example for two logarithmic CFTs to share the same 
basic set of conformal data but to differ in the detailed structure of their indecomposable representations.

In this paper, we consider the fusion algebras of the general logarithmic minimal models ${\cal LM}(p,p')$ and make explicit conjectures for their chiral fusion algebras. 
These fusion rules generalize our recent results~\cite{RP07} for critical percolation ${\cal LM}(2,3)$.
The generators of fusion are countably infinite in number but
the ensuing fusion rules are quasi-rational~\cite{Nahm94} in the sense that 
the fusion of a finite number of 
representations decomposes into a finite direct sum of representations. 
The conjectured fusion rules are commutative, associative and exhibit an $s\ell(2)$ 
structure at the level of characters. 
We make detailed comparisons of our fusion rules with the previous results of Gaberdiel 
and Kausch~\cite{GK96} for $p=1$ 
and with Eberle and Flohr~\cite{EF06} for $(p,p')=(2,5)$ coresponding to the logarithmic 
Yang-Lee model. 
In the latter case, we confirm that indecomposable representations of rank 3 arise 
as the result of certain lower-rank fusions. 
We also find that closure of a fundamental fusion algebra is achieved
without the introduction of indecomposable representations of rank higher than 3.
In general, the identity of the fundamental fusion algebra of ${\cal LM}(p,p')$
is a reducible yet indecomposable so-called Kac representation of rank 1.
The conjectured fusion rules are supported, within our lattice approach, 
by extensive numerical studies of  
the associated integrable lattice models. Details of our lattice findings and numerical results
will be presented elsewhere. The agreement of our results with previous results from the algebraic approach lends 
considerable support for the supposition that the logarithmic CFTs $c_{p,p'}$ and ${\cal LM}(p,p')$ should be identitifed.

\section{Representations of ${\cal LM}(p,p')$}

A logarithmic minimal model ${\cal LM}(p,p')$ is defined \cite{PRZ} for every coprime pair of
positive integers $p<p'$.
The model ${\cal LM}(p,p')$ has central charge
\be
 c\ =\  1-6\frac{(p'-p)^2}{pp'}
\label{c}
\ee
and conformal weights
\be
 \D_{r,s}\ =\ \frac{(rp'-sp)^{2}-(p'-p)^2}{4pp'},\hspace{1.2cm}r,s\in\mathbb{N}
\label{D}
\ee
The set of {\em distinct} values for the conformal weights is 
\bea
 S^{p,p'}\!\!&=&\!\!\{\Delta_{r,s};\ 1\leq r;\  1\leq s\leq p';\  0\leq rp'-sp\}\nn
  \!\!&=&\!\!\{\Delta_{r,s};\ 1\leq r\leq p;\  1\leq s;\  0\leq sp-rp'\}
\label{S}
\eea
This follows straightforwardly from the algebraic identities
\be
 \Delta_{r+kp,s+kp'}\ =\  \Delta_{-r+\ell p,-s+\ell p'}\ =\ \Delta_{r,s},\ \ \ \ \ 
  k,\ell\in\mathbb{Z}
\label{Dsymm}
\ee
and the fact that $\Delta_{r,s}\neq\Delta_{r+kp,s}$ and $\Delta_{r,s}\neq\Delta_{r,s+kp'}$ 
for $0\neq k\in\mathbb{Z}$.

\psset{unit=.85cm}
\begin{figure}[p]
\begin{center}
\begin{pspicture}(0,0)(7,11)
\psframe[linewidth=0pt,fillstyle=solid,fillcolor=lightlightblue](0,0)(7,11)
\psframe[linewidth=1pt,fillstyle=solid,fillcolor=lightblue](0,0)(0,1)
\multiput(0,0)(0,2){5}{\psframe[linewidth=0pt,fillstyle=solid,fillcolor=midblue](0,1)(7,2)}
\multirput(1,1)(1,0){6}{\pswedge[fillstyle=solid,fillcolor=red,linecolor=red](0,0){.25}{180}{270}}
\multirput(1,2)(1,0){6}{\pswedge[fillstyle=solid,fillcolor=red,linecolor=red](0,0){.25}{180}{270}}
\multirput(1,2)(0,2){5}{\pswedge[fillstyle=solid,fillcolor=red,linecolor=red](0,0){.25}{180}{270}}
\psgrid[gridlabels=0pt,subgriddiv=1]
\rput(.5,10.65){$\vdots$}\rput(1.5,10.65){$\vdots$}\rput(2.5,10.65){$\vdots$}\rput(3.5,10.65){$\vdots$}\rput(4.5,10.65){$\vdots$}\rput(5.5,10.65){$\vdots$}\rput(6.5,10.5){$\vvdots$}
\rput(.5,9.5){$\frac{63}8$}\rput(1.5,9.5){$\frac{35}8$}\rput(2.5,9.5){$\frac{15}8$}\rput(3.5,9.5){$\frac{3}8$}\rput(4.5,9.5){$-\frac 18$}\rput(5.5,9.5){$\frac{3}8$}\rput(6.5,9.5){$\cdots$}
\rput(.5,8.5){$6$}\rput(1.5,8.5){$3$}\rput(2.5,8.5){$1$}\rput(3.5,8.5){$0$}\rput(4.5,8.5){$0$}\rput(5.5,8.5){$1$}\rput(6.5,8.5){$\cdots$}
\rput(.5,7.5){$\frac{35}8$}\rput(1.5,7.5){$\frac {15}8$}\rput(2.5,7.5){$\frac 38$}\rput(3.5,7.5){$-\frac{1}8$}\rput(4.5,7.5){$\frac 38$}\rput(5.5,7.5){$\frac{15}8$}\rput(6.5,7.5){$\cdots$}
\rput(.5,6.5){$3$}\rput(1.5,6.5){$1$}\rput(2.5,6.5){$0$}\rput(3.5,6.5){$0$}\rput(4.5,6.5){$1$}\rput(5.5,6.5){$3$}\rput(6.5,6.5){$\cdots$}
\rput(.5,5.5){$\frac{15}8$}\rput(1.5,5.5){$\frac {3}{8}$}\rput(2.5,5.5){$-\frac 18$}\rput(3.5,5.5){$\frac{3}{8}$}\rput(4.5,5.5){$\frac {15}8$}\rput(5.5,5.5){$\frac{35}{8}$}\rput(6.5,5.5){$\cdots$}
\rput(.5,4.5){$1$}\rput(1.5,4.5){$0$}\rput(2.5,4.5){$0$}\rput(3.5,4.5){$1$}\rput(4.5,4.5){$3$}\rput(5.5,4.5){$6$}\rput(6.5,4.5){$\cdots$}
\rput(.5,3.5){$\frac 38$}\rput(1.5,3.5){$-\frac 18$}\rput(2.5,3.5){$\frac 38$}\rput(3.5,3.5){$\frac{15}8$}\rput(4.5,3.5){$\frac{35}8$}\rput(5.5,3.5){$\frac{63}8$}\rput(6.5,3.5){$\cdots$}
\rput(.5,2.5){$0$}\rput(1.5,2.5){$0$}\rput(2.5,2.5){$1$}\rput(3.5,2.5){$3$}\rput(4.5,2.5){$6$}\rput(5.5,2.5){$10$}\rput(6.5,2.5){$\cdots$}
\rput(.5,1.5){$-\frac 18$}\rput(1.5,1.5){$\frac 38$}\rput(2.5,1.5){$\frac{15}8$}\rput(3.5,1.5){$\frac{35}8$}\rput(4.5,1.5){$\frac{63}8$}\rput(5.5,1.5){$\frac{99}8$}\rput(6.5,1.5){$\cdots$}
\rput(.5,.5){$0$}\rput(1.5,.5){$1$}\rput(2.5,.5){$3$}\rput(3.5,.5){$6$}\rput(4.5,.5){$10$}\rput(5.5,.5){$15$}\rput(6.5,.5){$\cdots$}
{\color{blue}
\rput(.5,-.5){$1$}
\rput(1.5,-.5){$2$}
\rput(2.5,-.5){$3$}
\rput(3.5,-.5){$4$}
\rput(4.5,-.5){$5$}
\rput(5.5,-.5){$6$}
\rput(6.5,-.5){$r$}
\rput(-.5,.5){$1$}
\rput(-.5,1.5){$2$}
\rput(-.5,2.5){$3$}
\rput(-.5,3.5){$4$}
\rput(-.5,4.5){$5$}
\rput(-.5,5.5){$6$}
\rput(-.5,6.5){$7$}
\rput(-.5,7.5){$8$}
\rput(-.5,8.5){$9$}
\rput(-.5,9.5){$10$}
\rput(-.5,10.5){$s$}}
\end{pspicture}
\qquad\qquad%
\begin{pspicture}(0,0)(7,11)
\psframe[linewidth=0pt,fillstyle=solid,fillcolor=lightestblue](0,0)(7,11)
\psframe[linewidth=1pt,fillstyle=solid,fillcolor=lightblue](0,0)(1,2)
\psframe[linewidth=0pt,fillstyle=solid,fillcolor=lightlightblue](1,0)(2,11)
\psframe[linewidth=0pt,fillstyle=solid,fillcolor=lightlightblue](3,0)(4,11)
\psframe[linewidth=0pt,fillstyle=solid,fillcolor=lightlightblue](5,0)(6,11)
\psframe[linewidth=0pt,fillstyle=solid,fillcolor=lightlightblue](0,2)(7,3)
\psframe[linewidth=0pt,fillstyle=solid,fillcolor=lightlightblue](0,5)(7,6)
\psframe[linewidth=0pt,fillstyle=solid,fillcolor=lightlightblue](0,8)(7,9)
\multiput(0,0)(0,3){3}{\multiput(0,0)(2,0){3}{\psframe[linewidth=0pt,fillstyle=solid,fillcolor=midblue](1,2)(2,3)}}
\multirput(2,1)(2,0){3}{\pswedge[fillstyle=solid,fillcolor=red,linecolor=red](0,0){.25}{180}{270}}
\multirput(2,2)(2,0){3}{\pswedge[fillstyle=solid,fillcolor=red,linecolor=red](0,0){.25}{180}{270}}
\multirput(2,3)(2,0){3}{\pswedge[fillstyle=solid,fillcolor=red,linecolor=red](0,0){.25}{180}{270}}
\multirput(1,3)(0,3){3}{\pswedge[fillstyle=solid,fillcolor=red,linecolor=red](0,0){.25}{180}{270}}
\multirput(2,3)(0,3){3}{\pswedge[fillstyle=solid,fillcolor=red,linecolor=red](0,0){.25}{180}{270}}
\psgrid[gridlabels=0pt,subgriddiv=1]
\rput(.5,10.65){$\vdots$}\rput(1.5,10.65){$\vdots$}\rput(2.5,10.65){$\vdots$}\rput(3.5,10.65){$\vdots$}\rput(4.5,10.65){$\vdots$}\rput(5.5,10.65){$\vdots$}\rput(6.5,10.5){$\vvdots$}
\rput(.5,9.5){$12$}\rput(1.5,9.5){$\frac{65}8$}\rput(2.5,9.5){$5$}\rput(3.5,9.5){$\frac{21}8$}\rput(4.5,9.5){$1$}\rput(5.5,9.5){$\frac{1}8$}\rput(6.5,9.5){$\cdots$}
\rput(.5,8.5){$\frac{28}3$}\rput(1.5,8.5){$\frac{143}{24}$}\rput(2.5,8.5){$\frac{10}3$}\rput(3.5,8.5){$\frac{35}{24}$}\rput(4.5,8.5){$\frac 13$}\rput(5.5,8.5){$-\frac{1}{24}$}\rput(6.5,8.5){$\cdots$}
\rput(.5,7.5){$7$}\rput(1.5,7.5){$\frac {33}8$}\rput(2.5,7.5){$2$}\rput(3.5,7.5){$\frac{5}8$}\rput(4.5,7.5){$0$}\rput(5.5,7.5){$\frac{1}8$}\rput(6.5,7.5){$\cdots$}
\rput(.5,6.5){$5$}\rput(1.5,6.5){$\frac {21}8$}\rput(2.5,6.5){$1$}\rput(3.5,6.5){$\frac{1}8$}\rput(4.5,6.5){$0$}\rput(5.5,6.5){$\frac{5}8$}\rput(6.5,6.5){$\cdots$}
\rput(.5,5.5){$\frac{10}3$}\rput(1.5,5.5){$\frac {35}{24}$}\rput(2.5,5.5){$\frac 13$}\rput(3.5,5.5){$-\frac{1}{24}$}\rput(4.5,5.5){$\frac 13$}\rput(5.5,5.5){$\frac{35}{24}$}\rput(6.5,5.5){$\cdots$}
\rput(.5,4.5){$2$}\rput(1.5,4.5){$\frac 58$}\rput(2.5,4.5){$0$}\rput(3.5,4.5){$\frac{1}8$}\rput(4.5,4.5){$1$}\rput(5.5,4.5){$\frac{21}8$}\rput(6.5,4.5){$\cdots$}
\rput(.5,3.5){$1$}\rput(1.5,3.5){$\frac 18$}\rput(2.5,3.5){$0$}\rput(3.5,3.5){$\frac{5}8$}\rput(4.5,3.5){$2$}\rput(5.5,3.5){$\frac{33}8$}\rput(6.5,3.5){$\cdots$}
\rput(.5,2.5){$\frac 13$}\rput(1.5,2.5){$-\frac 1{24}$}\rput(2.5,2.5){$\frac 13$}\rput(3.5,2.5){$\frac{35}{24}$}\rput(4.5,2.5){$\frac{10}3$}\rput(5.5,2.5){$\frac{143}{24}$}\rput(6.5,2.5){$\cdots$}
\rput(.5,1.5){$0$}\rput(1.5,1.5){$\frac 18$}\rput(2.5,1.5){$1$}\rput(3.5,1.5){$\frac{21}8$}\rput(4.5,1.5){$5$}\rput(5.5,1.5){$\frac{65}8$}\rput(6.5,1.5){$\cdots$}
\rput(.5,.5){$0$}\rput(1.5,.5){$\frac 58$}\rput(2.5,.5){$2$}\rput(3.5,.5){$\frac{33}8$}\rput(4.5,.5){$7$}\rput(5.5,.5){$\frac{85}8$}\rput(6.5,.5){$\cdots$}
{\color{blue}
\rput(.5,-.5){$1$}
\rput(1.5,-.5){$2$}
\rput(2.5,-.5){$3$}
\rput(3.5,-.5){$4$}
\rput(4.5,-.5){$5$}
\rput(5.5,-.5){$6$}
\rput(6.5,-.5){$r$}
\rput(-.5,.5){$1$}
\rput(-.5,1.5){$2$}
\rput(-.5,2.5){$3$}
\rput(-.5,3.5){$4$}
\rput(-.5,4.5){$5$}
\rput(-.5,5.5){$6$}
\rput(-.5,6.5){$7$}
\rput(-.5,7.5){$8$}
\rput(-.5,8.5){$9$}
\rput(-.5,9.5){$10$}
\rput(-.5,10.5){$s$}}
\end{pspicture}
\mbox{}\vspace{.3in}\mbox{}
%
\begin{pspicture}(0,0)(7,11)
\psframe[linewidth=0pt,fillstyle=solid,fillcolor=lightestblue](0,0)(7,11)
\psframe[linewidth=1pt,fillstyle=solid,fillcolor=lightblue](0,0)(2,3)
\psframe[linewidth=0pt,fillstyle=solid,fillcolor=lightlightblue](2,0)(3,11)
\psframe[linewidth=0pt,fillstyle=solid,fillcolor=lightlightblue](5,0)(6,11)
\psframe[linewidth=0pt,fillstyle=solid,fillcolor=lightlightblue](0,3)(7,4)
\psframe[linewidth=0pt,fillstyle=solid,fillcolor=lightlightblue](0,7)(7,8)
\multiput(0,0)(0,4){2}{\multiput(0,0)(3,0){2}{\psframe[linewidth=0pt,fillstyle=solid,fillcolor=midblue](2,3)(3,4)}}
\multirput(3,1)(3,0){2}{\pswedge[fillstyle=solid,fillcolor=red,linecolor=red](0,0){.25}{180}{270}}
\multirput(3,2)(3,0){2}{\pswedge[fillstyle=solid,fillcolor=red,linecolor=red](0,0){.25}{180}{270}}
\multirput(3,3)(3,0){2}{\pswedge[fillstyle=solid,fillcolor=red,linecolor=red](0,0){.25}{180}{270}}
\multirput(3,4)(3,0){2}{\pswedge[fillstyle=solid,fillcolor=red,linecolor=red](0,0){.25}{180}{270}}
\multirput(1,4)(0,4){2}{\pswedge[fillstyle=solid,fillcolor=red,linecolor=red](0,0){.25}{180}{270}}
\multirput(2,4)(0,4){2}{\pswedge[fillstyle=solid,fillcolor=red,linecolor=red](0,0){.25}{180}{270}}
\multirput(3,4)(0,4){2}{\pswedge[fillstyle=solid,fillcolor=red,linecolor=red](0,0){.25}{180}{270}}
\psgrid[gridlabels=0pt,subgriddiv=1]
\rput(.5,10.65){$\vdots$}\rput(1.5,10.65){$\vdots$}\rput(2.5,10.65){$\vdots$}\rput(3.5,10.65){$\vdots$}\rput(4.5,10.65){$\vdots$}\rput(5.5,10.65){$\vdots$}\rput(6.5,10.5){$\vvdots$}
\rput(.5,9.5){$\frac{225}{16}$}\rput(1.5,9.5){$\frac{161}{16}$}\rput(2.5,9.5){$\frac{323}{48}$}\rput(3.5,9.5){$\frac{65}{16}$}\rput(4.5,9.5){$\frac{33}{16}$}\rput(5.5,9.5){$\frac{35}{48}$}\rput(6.5,9.5){$\cdots$}
\rput(.5,8.5){$11$}\rput(1.5,8.5){$\frac{15}{2}$}\rput(2.5,8.5){$\frac{14}3$}\rput(3.5,8.5){$\frac{5}{2}$}\rput(4.5,8.5){$1$}\rput(5.5,8.5){$\frac{1}{6}$}\rput(6.5,8.5){$\cdots$}
\rput(.5,7.5){$\frac{133}{16}$}\rput(1.5,7.5){$\frac {85}{16}$}\rput(2.5,7.5){$\frac{143}{48}$}\rput(3.5,7.5){$\frac{21}{16}$}\rput(4.5,7.5){$\frac 5{16}$}\rput(5.5,7.5){$-\frac{1}{48}$}\rput(6.5,7.5){$\cdots$}
\rput(.5,6.5){$6$}\rput(1.5,6.5){$\frac {7}2$}\rput(2.5,6.5){$\frac 53$}\rput(3.5,6.5){$\frac{1}2$}\rput(4.5,6.5){$0$}\rput(5.5,6.5){$\frac{1}6$}\rput(6.5,6.5){$\cdots$}
\rput(.5,5.5){$\frac{65}{16}$}\rput(1.5,5.5){$\frac {33}{16}$}\rput(2.5,5.5){$\frac {35}{48}$}\rput(3.5,5.5){$\frac{1}{16}$}\rput(4.5,5.5){$\frac 1{16}$}\rput(5.5,5.5){$\frac{35}{48}$}\rput(6.5,5.5){$\cdots$}
\rput(.5,4.5){$\frac 52$}\rput(1.5,4.5){$1$}\rput(2.5,4.5){$\frac 16$}\rput(3.5,4.5){$0$}\rput(4.5,4.5){$\frac 12$}\rput(5.5,4.5){$\frac{5}3$}\rput(6.5,4.5){$\cdots$}
\rput(.5,3.5){$\frac{21}{16}$}\rput(1.5,3.5){$\frac 5{16}$}\rput(2.5,3.5){$-\frac 1{48}$}\rput(3.5,3.5){$\frac{5}{16}$}\rput(4.5,3.5){$\frac{21}{16}$}\rput(5.5,3.5){$\frac{143}{48}$}\rput(6.5,3.5){$\cdots$}
\rput(.5,2.5){$\frac 12$}\rput(1.5,2.5){$0$}\rput(2.5,2.5){$\frac 16$}\rput(3.5,2.5){$1$}\rput(4.5,2.5){$\frac{5}2$}\rput(5.5,2.5){$\frac{14}{3}$}\rput(6.5,2.5){$\cdots$}
\rput(.5,1.5){$\frac 1{16}$}\rput(1.5,1.5){$\frac 1{16}$}\rput(2.5,1.5){$\frac{35}{48}$}\rput(3.5,1.5){$\frac{33}{16}$}\rput(4.5,1.5){$\frac{65}{16}$}\rput(5.5,1.5){$\frac{323}{48}$}\rput(6.5,1.5){$\cdots$}
\rput(.5,.5){$0$}\rput(1.5,.5){$\frac 12$}\rput(2.5,.5){$\frac 53$}\rput(3.5,.5){$\frac 72$}\rput(4.5,.5){$6$}\rput(5.5,.5){$\frac{55}6$}\rput(6.5,.5){$\cdots$}
{\color{blue}
\rput(.5,-.5){$1$}
\rput(1.5,-.5){$2$}
\rput(2.5,-.5){$3$}
\rput(3.5,-.5){$4$}
\rput(4.5,-.5){$5$}
\rput(5.5,-.5){$6$}
\rput(6.5,-.5){$r$}
\rput(-.5,.5){$1$}
\rput(-.5,1.5){$2$}
\rput(-.5,2.5){$3$}
\rput(-.5,3.5){$4$}
\rput(-.5,4.5){$5$}
\rput(-.5,5.5){$6$}
\rput(-.5,6.5){$7$}
\rput(-.5,7.5){$8$}
\rput(-.5,8.5){$9$}
\rput(-.5,9.5){$10$}
\rput(-.5,10.5){$s$}}
\end{pspicture}
\qquad\qquad%
\begin{pspicture}(0,0)(7,11)
\psframe[linewidth=0pt,fillstyle=solid,fillcolor=lightestblue](0,0)(7,11)
\psframe[linewidth=1pt,fillstyle=solid,fillcolor=lightblue](0,0)(1,4)
\psframe[linewidth=0pt,fillstyle=solid,fillcolor=lightlightblue](1,0)(2,11)
\psframe[linewidth=0pt,fillstyle=solid,fillcolor=lightlightblue](3,0)(4,11)
\psframe[linewidth=0pt,fillstyle=solid,fillcolor=lightlightblue](5,0)(6,11)
\psframe[linewidth=0pt,fillstyle=solid,fillcolor=lightlightblue](0,4)(7,5)
\psframe[linewidth=0pt,fillstyle=solid,fillcolor=lightlightblue](0,9)(7,10)
\multiput(0,0)(0,5){2}{\multiput(0,0)(2,0){3}{\psframe[linewidth=0pt,fillstyle=solid,fillcolor=midblue](1,4)(2,5)}}
\multirput(2,1)(2,0){3}{\pswedge[fillstyle=solid,fillcolor=red,linecolor=red](0,0){.25}{180}{270}}
\multirput(2,2)(2,0){3}{\pswedge[fillstyle=solid,fillcolor=red,linecolor=red](0,0){.25}{180}{270}}
\multirput(2,3)(2,0){3}{\pswedge[fillstyle=solid,fillcolor=red,linecolor=red](0,0){.25}{180}{270}}
\multirput(2,4)(2,0){3}{\pswedge[fillstyle=solid,fillcolor=red,linecolor=red](0,0){.25}{180}{270}}
\multirput(2,5)(2,0){3}{\pswedge[fillstyle=solid,fillcolor=red,linecolor=red](0,0){.25}{180}{270}}
\multirput(1,5)(0,5){2}{\pswedge[fillstyle=solid,fillcolor=red,linecolor=red](0,0){.25}{180}{270}}
\multirput(2,5)(0,5){2}{\pswedge[fillstyle=solid,fillcolor=red,linecolor=red](0,0){.25}{180}{270}}
\psgrid[gridlabels=0pt,subgriddiv=1]
\rput(.5,10.65){$\vdots$}\rput(1.5,10.65){$\vdots$}\rput(2.5,10.65){$\vdots$}\rput(3.5,10.65){$\vdots$}\rput(4.5,10.65){$\vdots$}\rput(5.5,10.65){$\vdots$}\rput(6.5,10.5){$\vvdots$}
\rput(.5,9.5){$\frac {27}5$}\rput(1.5,9.5){$\frac{91}{40}$}\rput(2.5,9.5){$\frac 25$}\rput(3.5,9.5){$-\frac{9}{40}$}\rput(4.5,9.5){$\frac 25$}\rput(5.5,9.5){$\frac{91}{40}$}\rput(6.5,9.5){$\cdots$}
\rput(.5,8.5){$4$}\rput(1.5,8.5){$\frac{11}{8}$}\rput(2.5,8.5){$0$}\rput(3.5,8.5){$-\frac{1}{8}$}\rput(4.5,8.5){$1$}\rput(5.5,8.5){$\frac{27}{8}$}\rput(6.5,8.5){$\cdots$}
\rput(.5,7.5){$\frac {14}5$}\rput(1.5,7.5){$\frac {27}{40}$}\rput(2.5,7.5){$-\frac 15$}\rput(3.5,7.5){$\frac{7}{40}$}\rput(4.5,7.5){$\frac 95$}\rput(5.5,7.5){$\frac{187}{40}$}\rput(6.5,7.5){$\cdots$}
\rput(.5,6.5){$\frac 95$}\rput(1.5,6.5){$\frac {7}{40}$}\rput(2.5,6.5){$-\frac 15$}\rput(3.5,6.5){$\frac{27}{40}$}\rput(4.5,6.5){$\frac {14}5$}\rput(5.5,6.5){$\frac{247}{40}$}\rput(6.5,6.5){$\cdots$}
\rput(.5,5.5){$1$}\rput(1.5,5.5){$-\frac {1}{8}$}\rput(2.5,5.5){$0$}\rput(3.5,5.5){$\frac{11}{8}$}\rput(4.5,5.5){$4$}\rput(5.5,5.5){$\frac{63}{8}$}\rput(6.5,5.5){$\cdots$}
\rput(.5,4.5){$\frac 25$}\rput(1.5,4.5){$-\frac 9{40}$}\rput(2.5,4.5){$\frac 25$}\rput(3.5,4.5){$\frac{91}{40}$}\rput(4.5,4.5){$\frac {27}5$}\rput(5.5,4.5){$\frac{391}{40}$}\rput(6.5,4.5){$\cdots$}
\rput(.5,3.5){$0$}\rput(1.5,3.5){$-\frac 18$}\rput(2.5,3.5){$1$}\rput(3.5,3.5){$\frac{27}8$}\rput(4.5,3.5){$7$}\rput(5.5,3.5){$\frac{95}8$}\rput(6.5,3.5){$\cdots$}
\rput(.5,2.5){$-\frac 15$}\rput(1.5,2.5){$\frac 7{40}$}\rput(2.5,2.5){$\frac 95$}\rput(3.5,2.5){$\frac{187}{40}$}\rput(4.5,2.5){$\frac{44}5$}\rput(5.5,2.5){$\frac{567}{40}$}\rput(6.5,2.5){$\cdots$}
\rput(.5,1.5){$-\frac 15$}\rput(1.5,1.5){$\frac {27}{40}$}\rput(2.5,1.5){$\frac{14}5$}\rput(3.5,1.5){$\frac{247}{40}$}\rput(4.5,1.5){$\frac{54}5$}\rput(5.5,1.5){$\frac{667}{40}$}\rput(6.5,1.5){$\cdots$}
\rput(.5,.5){$0$}\rput(1.5,.5){$\frac {11}8$}\rput(2.5,.5){$4$}\rput(3.5,.5){$\frac{63}8$}\rput(4.5,.5){$13$}\rput(5.5,.5){$\frac{155}8$}\rput(6.5,.5){$\cdots$}
{\color{blue}
\rput(.5,-.5){$1$}
\rput(1.5,-.5){$2$}
\rput(2.5,-.5){$3$}
\rput(3.5,-.5){$4$}
\rput(4.5,-.5){$5$}
\rput(5.5,-.5){$6$}
\rput(6.5,-.5){$r$}
\rput(-.5,.5){$1$}
\rput(-.5,1.5){$2$}
\rput(-.5,2.5){$3$}
\rput(-.5,3.5){$4$}
\rput(-.5,4.5){$5$}
\rput(-.5,5.5){$6$}
\rput(-.5,6.5){$7$}
\rput(-.5,7.5){$8$}
\rput(-.5,8.5){$9$}
\rput(-.5,9.5){$10$}
\rput(-.5,10.5){$s$}}
\end{pspicture}
\end{center}
\caption{Extended Kac tables of conformal weights $\Delta_{r,s}$ for critical dense polymers ${\cal LM}(1,2)$, critical percolation ${\cal LM}(2,3)$, the logarithmic Ising ${\cal LM}(3,4)$ and logarithmic Yang-Lee ${\cal LM}(2,5)$ models.
In general, the entries relate to distinct Kac representations even if the conformal weights coincide.  For a given model, an irreducible representation exists 
for each unique conformal weight appearing in the Kac table. The Kac representations which also happen to be irreducible representations are marked with a shaded quadrant in the top-right corner. These do not exhaust the distinct values of the conformal weights.
The periodicity $\Delta_{r,s}=\Delta_{r+p,s+p'}$ is made manifest by the shading of the rows and columns.}
\end{figure}
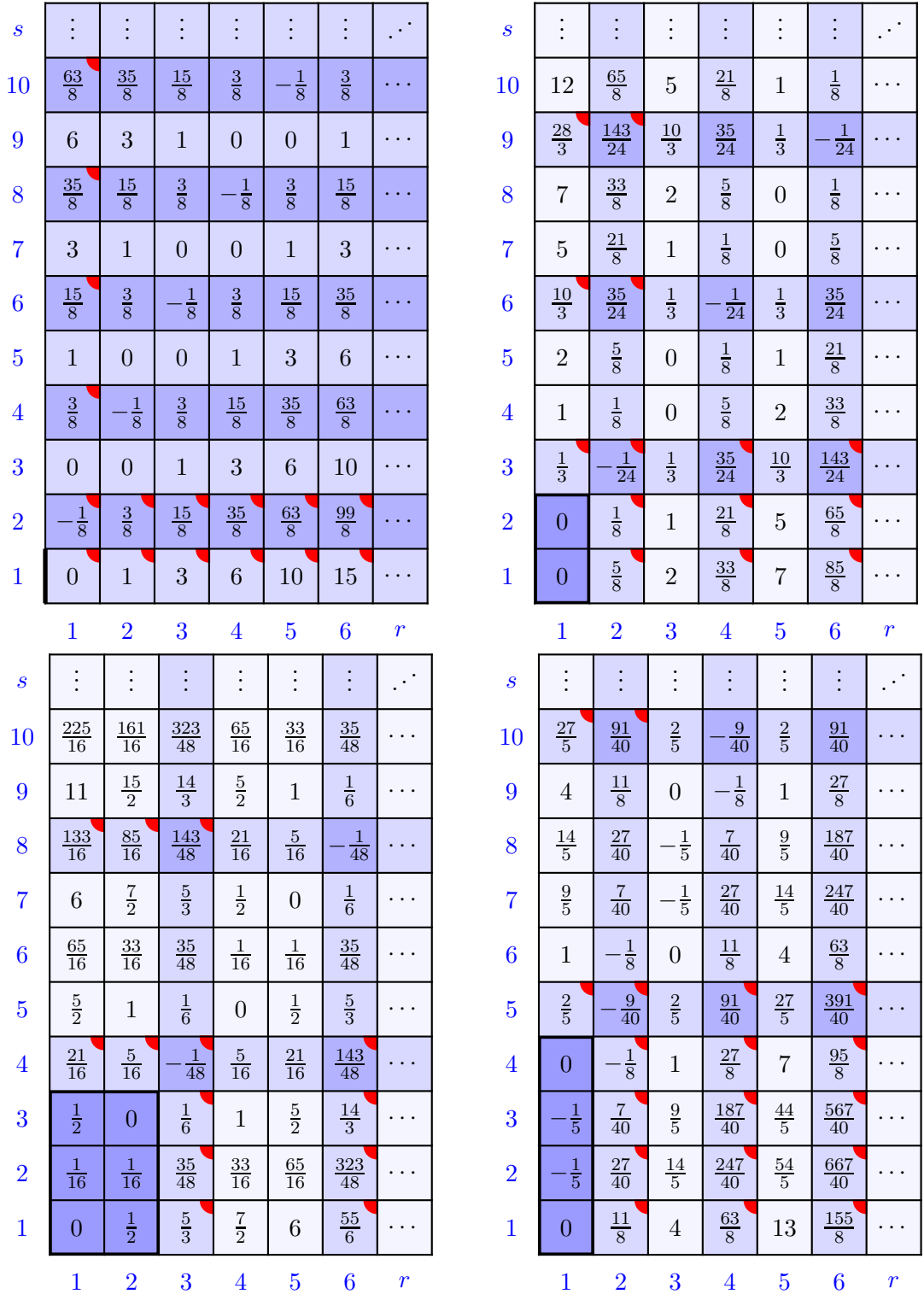
\psset{unit=1cm}

\subsection{Irreducible Characters}

There is a unique {\em irreducible} (highest-weight)
representation of conformal weight $\D_{r,s}$. It is denoted ${\cal V}(\D_{r,s})$ while
its character is denoted
\be
 \ch_{r,s}(q)\ =\ \chit[{\cal V}(\D_{r,s})](q)
\label{chchi}
\ee
As we will see, though, only a subset of these irreducible representations appear in
the present context while all the irreducible characters do. 
With $r_0=1,2,\ldots,p-1$, $s_0=1,2,\ldots,p'-1$ and $k\in\mathbb{N}-1$, 
these irreducible characters read \cite{FSZ}
\bea
 {\rm ch}_{r_0+kp,s_0}(q)\!\!&=&\!\!K_{2pp',(r_0+kp)p'-s_0p;k}(q)-K_{2pp',(r_0+kp)p'+s_0p;k}(q)\nn
 {\rm ch}_{r_0+(k+1)p,p'}(q)\!\!&=&\!\!
  \frac{1}{\eta(q)}\big(q^{(kp+r_0)^2p'/4p}-q^{((k+2)p-r_0)^2p'/4p}\big) \nn
 {\rm ch}_{(k+1)p,s_0}(q)\!\!&=&\!\!
  \frac{1}{\eta(q)}\big(q^{((k+1)p'-s_0)^2p/4p'}-q^{((k+1)p'+s_0)^2p/4p'}\big) \nn
 {\rm ch}_{(k+1)p,p'}(q)\!\!&=&\!\!
  \frac{1}{\eta(q)}\big(q^{k^2pp'/4}-q^{(k+2)^2pp'/4}\big)
\label{laq}
\eea
Here $K_{n,\nu;k}(q)$ is defined as
\be
 K_{n,\nu;k}(q)\ =\ \frac{1}{\eta(q)}\sum_{j\in\mathbb{Z}\setminus\{1,\ldots,k\}}q^{(\nu-jn)^2/2n}
\label{Kk}
\ee
while the Dedekind eta function is given by
\be
 \eta(q)\ =\ q^{1/24}\prod_{m=1}^\infty(1-q^m)
\label{eta}
\ee
It follows that for $k=0$, the first expression in (\ref{laq}) reduces to the well-known 
irreducible character 
\be
  {\rm ch}_{r_0,s_0}(q)\ =\ K_{2pp',r_0p'-s_0p}(q)-K_{2pp',r_0p'+s_0p}(q),\ \ \ \ \ 
  K_{n,\nu}(q)\ =\ \frac{1}{\eta(q)}\sum_{j\in\mathbb{Z}}q^{(\nu+jn)^2/2n}
\label{r0s0}
\ee

\subsection{Kac Representations}

{}From the lattice, a representation $(r,s)$, which we call a {\em Kac representation}, 
arises for {\em every} pair of integer Kac labels $r,s$ in the first quadrant
of the infinitely extended Kac table, see Figure~1. 
This relaxes the constraint $r=1,2,\ldots,p$ considered in
\cite{PRZ}. The lattice description of the full set of Kac representations
will be discussed in detail elsewhere.
The conformal character of the Kac representation $(r,s)$ is given by
\be
 \chit_{r,s}(q)\ =\ \frac{q^{\frac{1-c}{24}+\D_{r,s}}}{\eta(q)}\big(1-q^{rs}\big)
  \ =\ \frac{1}{\eta(q)}\big(q^{(rp'-sp)^2/4pp'}-q^{(rp'+sp)^2/4pp'}\big)
\label{chikac}
\ee
corresponding to the Virasoro character of the quotient module $V_{r,s}/V_{r,-s}$ of the two
highest-weight Verma modules $V_{r,s}=V_{\D_{r,s}}$ and $V_{r,-s}=V_{\D_{r,-s}}$.
A priori, a Kac representation can be either irreducible or reducible.
We will only characterize the Kac representations appearing in the
fusion algebras to be discussed in the present work.

Among these are the {\em irreducible} Kac representations 
\be
 \{(r,kp'),(kp,s);\ r=1,2,\ldots,p;\ s=1,2,\ldots,p';\ k\in\mathbb{N}\}
\label{Kacirr}
\ee
Since their characters
all correspond to irreducible Virasoro characters, these Kac representations must indeed
themselves be irreducible. The set (\ref{Kacirr}) constitutes an 
exhaustive list of irreducible Kac representations.
Two Kac representations are naturally identified if they have identical
conformal weights and are both irreducible. The relations
\be
 (kp,p')\ =\ (p,kp')
\label{idirr}
\ee
are the only such identifications. 
More general relations are considered in (\ref{kkexp}) and (\ref{RequalR}).
For now, we simply point out that two Kac characters (\ref{chikac}) 
are equal $\chit_{r,s}(q)=\chit_{r',s'}(q)$ if and only if $(r',s')=(r,s)$ or $(r',s')=(sp/p',rp'/p)$.
That is, the only equalities between Kac characters are of the form $\chit_{kp,k'p'}(q)=\chit_{k'p,kp'}(q)$.
According to (\ref{RequalR}), a similar equality applies to the Kac representations themselves:
$(kp,k'p')=(k'p,kp')$.

Somewhat redundantly, we also encounter {\em fully reducible} Kac representations
\be
 \{(kp,k'p');\ k,k'\in\mathbb{N}+1\}
\label{KacFully}
\ee
Since they decompose into direct sums of irreducible representations, cf.$\,$(\ref{kkexp}),
they only enter the fusion analysis in intermediate steps.

Finally, the Kac representations 
\be
 \{(r_0,s_0);\ r_0=1,2,\ldots,p-1;\ s_0=1,2,\ldots,p'-1\}
\label{KacRank1}
\ee
are {\em reducible yet indecomposable} representations of rank 1. 
It is noted that these representations occupy the lower-left
corner of the infinitely extended Kac table
corresponding to the Kac table of the rational 
cousin of ${\cal LM}(p,p')$ --- the minimal model characterized by $p,p'$. 
One may view these reducible yet indecomposable
representations as `logarithmic replacements' of the irreducible representations associated
to the rational Kac table.
As discussed in 
\cite{EF06,RP07} in the case of critical percolation ${\cal LM}(2,3)$, 
the representations (\ref{KacRank1}) can be viewed also as subrepresentations
of certain indecomposable representations of rank 2. 
In the general case, these indecomposable representations are denoted $\R_{r_0,p'}^{0,p'-s_0}$ and
$\R_{p,s_0}^{p-r_0,0}$ in the following. These and all other indecomposable
representations of higher rank appearing in our fusion analysis will be discussed below.

The indecomposable representations of higher rank 
may be described in terms of Kac representations and their characters.
We therefore list the decompositions of the relevant Kac characters in terms of
irreducible characters 
\bea
 \chit_{pk-r_0,s_0}(q)\!\!&=&\!\! \ch_{pk-r_0,s_0}(q)+\ch_{pk+r_0,s_0}(q)\nn
 \chit_{pk-r_0,p'}(q)\!\!&=&\!\! \ch_{pk-r_0,p'}(q)+\big(1-\delta_{k,1}\big)\ch_{pk+r_0,p'}(q)\nn
 \chit_{pk+r_0,p'+s_0}(q)\!\!&=&\!\! \ch_{p(k-1)+r_0,s_0}(q)+\ch_{pk+r_0,p'-s_0}(q)
  +\ch_{p(k+1)-r_0,s_0}(q)+\ch_{p(k+1)+r_0,s_0}(q)\nn
 \!\!&+&\!\!\ch_{p(k+2)-r_0,p'-s_0}(q)+\ch_{p(k+3)-r_0,s_0}(q)  \nn
 \chit_{r_0,kp'-s_0}(q)\!\!&=&\!\!\ch_{r_0,p'k-s_0}(q)+\ch_{r_0,p'k+s_0}(q)\nn
 \chit_{p,p'k-s_0}(q)\!\!&=&\!\! \ch_{p,p'k-s_0}(q)+\big(1-\delta_{k,1}\big)\ch_{p,p'k+s_0}(q)\nn
 \chit_{p+r_0,p'k+s_0}(q)\!\!&=&\!\! \ch_{r_0,p'(k-1)+s_0}(q)+\ch_{p-r_0,p'k+s_0}(q)
  +\ch_{r_0,p'(k+1)-s_0}(q)+\ch_{r_0,p'(k+1)+s_0}(q)\nn
 \!\!&+&\!\!\ch_{p-r_0,p'(k+2)-s_0}(q)+\ch_{r_0,p'(k+3)-s_0}(q) 
\label{chitch}
\eea
where $r_0=1,2,\ldots,p-1$ and $s_0=1,2,\ldots,p'-1$ whereas $k\in\mathbb{N}$. 
The decomposition of a general Kac character $\chit_{r,s}(q)$ into irreducible characters
is discussed in the appendix of \cite{PRZ}.

\subsection{Indecomposable Representations of Rank 2 or 3}

{}From the lattice analysis, we infer that the logarithmic minimal model ${\cal LM}(p,p')$
contains indecomposable representations of rank 2 and for $p>1$ also indecomposable
representations of rank 3.
For $a,r_0=1,2,\ldots,p-1$ and $b,s_0=1,2,\ldots,p'-1$ as well as $k\in\mathbb{N}$, 
the representations denoted by $\R_{pk,s_0}^{a,0}$, $\R_{pk,p'}^{a,0}$, 
$\R_{r_0,p'k}^{0,b}$ and $\R_{p,p'k}^{0,b}$ are indecomposable
representations of rank 2, while $\R_{pk,p'}^{a,b}$ is an indecomposable representation of rank 3.
Their characters read
\bea
 \chit[\R_{pk,s_0}^{a,0}](q)\!\!&=&\!\!\chit_{pk-a,s_0}(q)+\chit_{pk+a,s_0}(q)
  \ =\ \ch_{pk-a,s_0}(q)+2\ch_{pk+a,s_0}(q)+\ch_{p(k+2)-a,s_0}(q)\nn
 \chit[\R_{pk,p'}^{a,0}](q)\!\!&=&\!\!\chit_{pk-a,p'}(q)+\chit_{pk+a,p'}(q)
   \ =\ \big(1-\delta_{k,1}\big)\ch_{pk-a,p'}(q)+2\ch_{pk+a,p'}(q)+\ch_{p(k+2)-a,p'}(q)\nn
 \chit[\R_{r_0,p'k}^{0,b}](q)\!\!&=&\!\!\chit_{r_0,p'k-b}(q)+\chit_{r_0,p'k+b}(q)
    \ =\ \ch_{r_0,p'k-b}(q)+2\ch_{r_0,p'k+b}(q)+\ch_{r_0,p'(k+2)-b}(q)\nn
 \chit[\R_{p,p'k}^{0,b}](q)\!\!&=&\!\!\chit_{a,3k-b}(q)+\chit_{a,3k+b}(q)
  \ =\ \big(1-\delta_{k,1}\big)\ch_{p,p'k-b}(q)+2\ch_{p,p'k+b}(q)+\ch_{p,p'(k+2)-b}(q)\nn
 \chit[\R_{pk,p'}^{a,b}](q)\!\!&=&\!\!\chit_{pk-a,p'-b}(q)+\chit_{pk-a,p'+b}(q)
  +\chit_{pk+a,p'-b}(q)+\chit_{pk+a,p'+b}(q)\nn
 \!\!&=&\!\!  \big(1-\delta_{k,1}\big)\ch_{p(k-1)-a,b}(q)+2\ch_{p(k-1)+a,b}(q)
   +2\big(1-\delta_{k,1}\big)\ch_{pk-a,p'-b}(q)\nn
 \!\!&+&\!\!4\ch_{pk+a,p'-b}(q)+\big(2-\delta_{k,1}\big)\ch_{p(k+1)-a,b}(q)
  +2\ch_{p(k+1)+a,b}(q)\nn
 \!\!&+&\!\!2\ch_{p(k+2)-a,p'-b}(q)+\ch_{p(k+3)-a,b}(q)\nn
 \!\!&=&\!\!  \big(1-\delta_{k,1}\big)\ch_{a,p'(k-1)-b}(q)+2\ch_{a,p'(k-1)+b}(q)
   +2\big(1-\delta_{k,1}\big)\ch_{p-a,p'k-b}(q)\nn
 \!\!&+&\!\!4\ch_{p-a,p'k+b}(q)+\big(2-\delta_{k,1}\big)\ch_{a,p'(k+1)-b}(q)
  +2\ch_{a,p'(k+1)+b}(q)\nn
 \!\!&+&\!\!2\ch_{p-a,p'(k+2)-b}(q)+\ch_{a,p'(k+3)-b}(q)
\label{chiR}
\eea
indicating that one may consider these indecomposable representations
as `indecomposable combinations' of Kac representations. The participating
Kac representations are of course the ones whose characters appear
in (\ref{chiR}). In the case of the indecomposable representation $\R_{pk,s}^{a,0}$
(or $\R_{r,p'k}^{0,b}$) of rank 2, our lattice analysis indicates 
that a Jordan cell is formed between every state in $\ch_{pk+a,s}(q)$
(or $\ch_{r,p'k+b}(q)$) and its partner state in the second copy of
$\ch_{pk+a,s}(q)$ (or $\ch_{r,p'k+b}(q)$), and nowhere else.
In the case of the indecomposable representation $\R_{pk,p'}^{a,b}$ of rank 3,
our lattice analysis indicates that for every quartet of matching states in the four 
copies of $\ch_{pk+a,p'-b}(q)=\ch_{p-a,p'k+b}(q)$, 
a rank-3 Jordan cell is formed along with a single state. It likewise
appears that a Jordan cell of rank 2 is formed between every pair of matching
states in the irreducible components with multiplicity 2.

The notation $\R_{r,s}^{a,b}$ is meant to reflect simple properties of the
higher-rank indecomposable representations. The pair of lower indices
thus refers to a `symmetry point' in the Kac table around which an indecomposable combination 
of Kac representations is located. The pair of upper indices indicates 
the distribution of these representations of which there are either two
(if $a=0$ or $b=0$) or four (if $a,b\neq0$). Their locations correspond to 
endpoints or corners, respectively, of a line segment or a rectangle with centre
at $(r,s)$. This structure is encoded neatly in the character expressions (\ref{chiR}).
Setting 
\be
 \R_{pk,p'k'}^{0,0}\ =\ (pk,p'k')
\label{R00}
\ee
the representation $\R_{pk,p'k'}^{a,b}$ thus has rank $d+1=1,2,3$ if, in the Kac table, it 
corresponds to the corners of a $d$-dimensional rectangle with centre at $(pk,p'k')$, width $2a$
and height $2b$.

\section{Fundamental Component Fusion Algebra of ${\cal LM}(p,p')$}
\label{onelegged}

The {\em fundamental} fusion algebra 
\be
 \big\langle(2,1), (1,2)\big\rangle_{p,p'}
\label{fund}
\ee 
is defined as the fusion algebra
generated by the {\em fundamental Kac representations} $(2,1)$ and $(1,2)$.
It follows from the lattice description that the fundamental fusion algebra
is both associative and commutative.
It also follows from the lattice that the fusion algebra may be described
by separating the representations into a horizontal and a vertical part.
Before discussing implications of this, we examine the two directions separately.
That is, we initially consider the horizontal fusion algebra
\be
 \big\langle(2,1)\big\rangle_{p,p'} \ =\ 
   \big\langle(r_0,1), (pk,1), \R_{pk,1}^{a,0};\ r_0,a=1,2,\ldots,p-1;\ k\in\mathbb{N} \big\rangle_{p,p'}
\label{hor}
\ee 
and the vertical fusion algebra
\be
 \big\langle(1,2)\big\rangle_{p,p'}\ =\ 
   \big\langle(1,s_0), (1,p'k), \R_{1,p'k}^{0,b};\ s_0,b=1,2,\ldots,p'-1;\ k\in\mathbb{N} \big\rangle_{p,p'}
\label{ver}
\ee
in their own right.
By abbreviating the set of representations
$\{(r_0,1), (pk,1), \R_{pk,1}^{a,0}\}$ by $\{(r_0), (pk), \R_{pk}^a\}$
and similarly $\{(1,s_0), (1,p'k), \R_{1,p'k}^{0,b}\}$ by $\{(s_0), (p'k), \R_{p'k}^b\}$,
this can be done in one go.
Despite the following choice of dummy variables in these abbreviations, 
this notation can represent either direction, and the ensuing fusion algebra
\be
 \big\langle(2)\big\rangle_p\ =\ 
  \big\langle(r_0), (pk)=\R_{pk}^0, \R_{pk}^a;\ r_0,a=1,2,\ldots,p-1;\ k\in\mathbb{N} \big\rangle_p
\label{fund2}
\ee
will henceforth be referred to as the fundamental {\em component} fusion algebra of {\em order} $p$.
To unify the notation, we have introduced 
\be
 \R_{pk}^0\ =\ (pk)
\ee
and will use the notation
\be
 (-r)\ \equiv\ -(r),\hspace{1cm} \R_{-r}^{a}\ \equiv\ -\R_{r}^{a}
\ee
implying, in particular, that $(0)\equiv\R_{0}^{a}\equiv0$.
Following \cite{RP07}, we also introduce the Kronecker delta combinations
\bea
 \dnn\!\!&=&\!\!  2-\delta_{j,|n-n'|}-\delta_{j,n+n'}  \nn
 \ddnn\!\!&=&\!\!   4-3\delta_{j,|n-n'|-1}-2\delta_{j,|n-n'|}-\delta_{j,|n-n'|+1}
   -\delta_{j,n+n'-1}-2\delta_{j,n+n'}-3\delta_{j,n+n'+1}
   \nn
 \dddnn\!\!&=&\!\!8-7\delta_{j,|n-n'|-2}-6\delta_{j,|n-n'|-1}-4\delta_{j,|n-n'|}-2\delta_{j,|n-n'|+1}
  -\delta_{j,|n-n'|+2}\nn
  \!\!&-&\!\!\delta_{j,n+n'-2}-2\delta_{j,n+n'-1}-4\delta_{j,n+n'}-6\delta_{j,n+n'+1}-7\delta_{j,n+n'+2}
\label{d24}
\eea

For $a,a',r_0,r'_0=1,2,\ldots,p-1$, our conjecture for the fusion rules of the fundamental component 
fusion algebra of order $p$ is
\bea
 (r_0)\otimes\R_{pn}^0\!\!&=&\!\! \bigoplus_{i=0}^{\lfloor\frac{r_0-1}{2}\rfloor}\ \R_{pn}^{r_0-1-2i}    \nn
 \R_{pn}^{0}\otimes\R_{pn'}^{0}\!\!&=&\!\! \bigoplus_{j=|n-n'|+1,\ \!{\rm by}\ \!2}^{n+n'-1}
   \Big\{\bigoplus_{i=0}^{\lfloor\frac{p-1}{2}\rfloor}\R_{pj}^{p-1-2i}\Big\}\nn
 \R_{pn}^0\otimes\R_{pn'}^{a'}\!\!&=&\!\!
   \Big(\bigoplus_{j=|n-n'|,\ \!{\rm by}\ \!2}^{n+n'}\dnn\Big\{\bigoplus_{i=0}^{\lfloor\frac{a'-1}{2}\rfloor}
     \R_{pj}^{a'-1-2i}\Big\}\Big)
  \oplus\Big(\bigoplus_{j=|n-n'|+1,\ \!{\rm by}\ \!2}^{n+n'-1}2\ \! \Big\{
    \bigoplus_{i=0}^{\lfloor\frac{p-a'-1}{2}\rfloor}\R_{pj}^{p-a'-1-2i}\Big\}\Big)\nn
\eea
where for $r_0+r'_0,r_0+a,a+a'\leq p$
\bea
 (r_0)\otimes(r'_0)\!\!&=&\!\! \bigoplus_{j=|r_0-r'_0|+1,\ \!{\rm by}\ \!2}^{r_0+r'_0-1}(j)   \nn
 (r_0)\otimes\R_{pn}^a\!\!&=&\!\!  
  \Big\{\bigoplus_{i=0}^{\min\{r_0-1,\lfloor\frac{r_0+a-1}{2}\rfloor\}}\R_{pn}^{r_0+a-1-2i}\Big\}
   \oplus\Big\{\bigoplus_{i=0}^{\lfloor\frac{r_0-a-1}{2}\rfloor}\R_{pn}^{r_0-a-1-2i}\Big\}   \nn
 \R_{pn}^a\otimes\R_{pn'}^{a'} \!\!&=&\!\!
   \Big(
     \bigoplus_{j=|n-n'|,\ \!{\rm by}\ \!2}^{n+n'}\dnn
      \Big\{\big(\bigoplus_{i=0}^{\lfloor\frac{|a-a'|-1}{2}\rfloor}\R_{pj}^{|a-a'|-1-2i}\Big)     
         \oplus\Big(\bigoplus_{i=0}^{\lfloor\frac{a+a'-1}{2}\rfloor}\R_{pj}^{a+a'-1-2i}\Big)\Big\}\Big)\nn
 \!\!&\oplus&\!\!
   \Big(\bigoplus_{j=|n-n'|+1,\ \!{\rm by}\ \!2}^{n+n'-1}
    2\ \! \Big\{\Big(\bigoplus_{i=0}^{\lfloor\frac{p-|a-a'|-1}{2}\rfloor}\R_{pj}^{p-|a-a'|-1-2i}\Big)
    \oplus\Big(\bigoplus_{i=0}^{\lfloor\frac{p-a-a'-1}{2}\rfloor}\R_{pj}^{p-a-a'-1-2i}\Big)\Big\}\Big)\nn
\eea
while for $r_0+r'_0,r_0+a,a+a'>p$
\bea
 (r_0)\otimes(r'_0)\!\!&=&\!\! 
   \Big(\bigoplus_{j=|r_0-r'_0|+1,\ \!{\rm by}\ \!2}^{2p-r_0-r'_0-1}(j)\Big)
  \oplus\Big\{ \bigoplus_{i=0}^{\lfloor\frac{r_0+r'_0-p-1}{2}\rfloor} \R_{p}^{r_0+r'_0-p-1-2i}\Big\}   \nn
 (r_0)\otimes\R_{pn}^a\!\!&=&\!\! \Big\{
   \bigoplus_{i=0}^{\lfloor\frac{r_0+a-p-1}{2}\rfloor}\Big(\R_{pn-p}^{r_0+a-p-1-2i}
     \oplus\R_{pn+p}^{r_0+a-p-1-2i}\Big)     \Big\}\nn
 \!\!&\oplus&\!\!
   \Big\{\bigoplus_{i=0}^{\min\{p-a-1,\lfloor\frac{2p-r_0-a-1}{2}\rfloor\}}\R_{pn}^{2p-r_0-a-1-2i}\Big\}
    \oplus\Big\{\bigoplus_{i=0}^{\lfloor\frac{r_0-a-1}{2}\rfloor}\R_{pn}^{r_0-a-1-2i}\Big\}\nn
 \R_{pn}^a\otimes\R_{pn'}^{a'}\!\!&=&\!\!
  \Big(\bigoplus_{j=|n-n'|-1,\ \!{\rm by}\ \!2}^{n+n'+1}\ddnn
    \Big\{\bigoplus_{i=0}^{\lfloor\frac{a+a'-p-1}{2}\rfloor}\R_{pj}^{a+a'-p-1-2i}\Big\}\Big)\nn
 \!\!&\oplus&\!\!
   \Big(
     \bigoplus_{j=|n-n'|,\ \!{\rm by}\ \!2}^{n+n'}\dnn
      \Big\{\Big(\bigoplus_{i=0}^{\lfloor\frac{|a-a'|-1}{2}\rfloor}\R_{pj}^{|a-a'|-1-2i}\Big)     
         \oplus\Big(\bigoplus_{i=0}^{\lfloor\frac{2p-a-a'-1}{2}\rfloor}\R_{pj}^{2p-a-a'-1-2i}\Big)\Big\}\Big)\nn
 \!\!&\oplus&\!\!
   \Big(\bigoplus_{j=|n-n'|+1,\ \!{\rm by}\ \!2}^{n+n'-1}
    2\ \! \Big\{\bigoplus_{i=0}^{p-\max\{a,a'\}-1}\R_{pj}^{p-|a-a'|-1-2i}\Big\}\Big)
\label{aap}
\eea
These expressions can of course be combined into more condensed but less transparent expressions, 
which are not included here. Particular subalgebras, though, 
of these fundamental component fusion algebras
are written out and simplified in Appendix \ref{appEX}.

\section{Fundamental Fusion Algebra of ${\cal LM}(p,p')$}

We find that closure of the fundamental fusion algebra (\ref{fund})
requires the inclusion of the many various representations discussed above
\be
  \big\langle(2,1), (1,2)\big\rangle_{p,p'}\ =\ \big\langle(r_0,s_0), (pk,s_0), (r_0,p'k), (pk,p'),
   \R_{pk,s_0}^{a,0}, \R_{pk,p'}^{a,0}, \R_{r_0,p'k}^{0,b}, \R_{p,p'k}^{0,b}  
  \R_{pk,p'}^{a,b}\big\rangle_{p,p'}
\label{A2112}
\ee
where $r_0,a=1,2,\ldots,p-1$ and $s_0,b=1,2,\ldots,p'-1$ whereas 
$k\in\mathbb{N}$. According to (\ref{idirr}) and (\ref{RequalR}) below, 
$(pk,p')=(p,p'k)$ and $\R_{pk,p'}^{a,b}=\R_{p,p'k}^{a,b}$, restoring the apparent lack of symmetry
in the list (\ref{A2112}).

In the following, we will discuss how this fundamental fusion algebra may be obtained by combining
two fundamental component fusion algebras of order $p$ and $p'$, respectively,
and present explicit examples and comparisons with the literature.
To compactify the fusion rules, we will use the notation
\be
 (r,-s)\ \equiv\ (-r,s)\ \equiv\ -(r,s),\hspace{1cm}\R_{-r,s}^{a,b}\ \equiv\ \R_{r,-s}^{a,b}\ 
   \equiv\ -\R_{r,s}^{a,b}
\ee
implying, in particular, that $(0,s)\equiv(r,0)\equiv\R_{0,s}^{a,b}\equiv\R_{r,0}^{a,b}\equiv0$.

\subsection{Decomposition into Horizontal and Vertical Fusion}

With $a=0,1,\ldots,p-1$ and $b=0,1,\ldots,p'-1$, we introduce the representations
\be
 \R_{pk,p'k'}^{a,b}\ =\ \R_{pk,1}^{a,0}\otimes \R_{1,p'k'}^{0,b}
\label{kk}
\ee
thus defined as simple fusions of `a horizontal and a vertical representation'.
Combining these with the associativity and commutativity of the fusion rules
results in a separation of fusion itself into a horizontal and a vertical part. 
We illustrate this with a general but somewhat formal evaluation. Letting 
$A_{r,s}=\bar{a}_{r,1}\otimes\ a_{1,s}$, $B_{r',s'}=\bar{b}_{r',1}\otimes\ b_{1,s'}$,
$\bar{a}_{r,1}\otimes\bar{b}_{r',1}=\bigoplus_{r''}\bar{c}_{r'',1}$ and
$a_{1,s}\otimes b_{1,s'}=\bigoplus_{s''}c_{1,s''}$, our fusion prescription yields
\bea
 A_{r,s}\otimes B_{r',s'}\!\!&=&\!\!\Big(\bar{a}_{r,1}\otimes a_{1,s}\Big)\otimes
  \Big(\bar{b}_{r',1}\otimes b_{1,s'}\Big)
   \ =\ \Big(\bar{a}_{r,1}\otimes\bar{b}_{r',1}\Big)\otimes
  \Big(a_{1,s}\otimes b_{1,s'}\Big)\nn
 \!\!&=&\!\!\Big(\bigoplus_{r''}\bar{c}_{r'',1}\Big)\otimes\Big(\bigoplus_{s''}c_{1,s''}\Big)
  \ =\ \bigoplus_{r'',s''}C_{r'',s''}
\label{rs}
\eea
where $C_{r'',s''}=\bar{c}_{r'',1}\otimes c_{1,s''}$.
As already indicated, this way of evaluating the fusion of two representations 
follows from the lattice description and will be used repeatedly in the following.

\subsection{Decompositions of Representations}

The representations defined in (\ref{kk}) where $a=0,1,\ldots,p-1$ and $b=0,1,\ldots,p'-1$
can be decomposed as
\be
 \R_{pk,p'k'}^{a,b}\ =\ \bigoplus_{j=|k-k'|+1,\ \!{\rm by}\ \!2}^{k+k'-1}\R_{pj,p'}^{a,b}
  \ =\ \bigoplus_{j=|k-k'|+1,\ \!{\rm by}\ \!2}^{k+k'-1}\R_{p,p'j}^{a,b}
\label{kkexp}
\ee
with
\be
 \R_{pk,p'k'}^{a,b}\ =\ \R_{pk',p'k}^{a,b}
\label{RequalR}
\ee 
as special identifications extending the set (\ref{idirr}). 
For critical percolation ${\cal LM}(2,3)$, the decompositions (\ref{kkexp}) and identifications 
(\ref{RequalR}) already appeared in \cite{RP07}, though without proof.
Here we establish, by induction in $k$ and $k'$,
that (\ref{kkexp}) is a consequence of our fusion rules for general ${\cal LM}(p,p')$.
The induction start corresponds to 
\be
 \R_{pm,p'}^{a,b}\ =\ \R_{p,p'm}^{a,b}
\label{start}
\ee 
for $m\in\mathbb{N}$, while the induction step amounts to establishing that if (\ref{kkexp}) is valid for 
$k=1,2,\ldots,n$ and $k'=1,2,\ldots,n'$, then (\ref{kkexp}) is valid also for 
$k=n+1$ and independently for $k'=n'+1$.
It is noted that 
the second equality in (\ref{kkexp}) is an immediate consequence of the induction start.

To establish (\ref{start}), we use (\ref{idirr}) and consider
\be
 (r,s)\otimes(pm,p')\ =\ (r,s)\otimes(p,p'm),\hspace{1.5cm}r=1,2,\ldots,p;\ s=1,2,\ldots,p'; \ m\in\mathbb{N}
\ee
Following the fusion prescription (\ref{rs}), the left side reads
\bea
 \big((r,1)\otimes(pm,1)\big)\otimes\big((1,s)\otimes(1,p')\big)\!\!&=&\!\!
  \Big(\bigoplus_{i=0}^{\lfloor\frac{r-1}{2}\rfloor}\R_{pm,1}^{r-1-2i,0}\Big)
  \otimes\Big(\bigoplus_{i'=0}^{\lfloor\frac{s-1}{2}\rfloor}\R_{1,p'}^{0,s-1-2i'}\Big)\nn
 \!\!&=&\!\!\bigoplus_{i=0}^{\lfloor\frac{r-1}{2}\rfloor}\bigoplus_{i'=0}^{\lfloor\frac{s-1}{2}\rfloor}
  \R_{pm,p'}^{r-1-2i,s-1-2i'}
\eea
while the right side reads
\be
 \big((r,1)\otimes(p,1)\big)\otimes\big((1,s)\otimes(1,p'm)\big)\ =\ 
   \bigoplus_{i=0}^{\lfloor\frac{r-1}{2}\rfloor}\bigoplus_{i'=0}^{\lfloor\frac{s-1}{2}\rfloor}
    \R_{p,p'm}^{r-1-2i,s-1-2i'}
\ee
yielding
\bea
 \R_{p,p'm}^{r-1,s-1}\ominus\R_{pm,p'}^{r-1,s-1}\!\!&=&\!\!
  \bigoplus_{i=1}^{\lfloor\frac{r-1}{2}\rfloor}\bigoplus_{i'=0}^{\lfloor\frac{s-1}{2}\rfloor}
   \big(\R_{pm,p'}^{r-1-2i,s-1-2i'}\ominus\R_{p,p'm}^{r-1-2i,s-1-2i'}\big)\nn
 \!\!&\oplus&\!\!\bigoplus_{i=0}^{\lfloor\frac{r-1}{2}\rfloor}\bigoplus_{i'=1}^{\lfloor\frac{s-1}{2}\rfloor}
   \big(\R_{pm,p'}^{r-1-2i,s-1-2i'}\ominus\R_{p,p'm}^{r-1-2i,s-1-2i'}\big)
\label{starteq}
\eea
Here and in the following, the notation $A\ominus B=C$ is equivalent to the direct-sum
decomposition $A=B\oplus C$. The induction start (\ref{start})
now follows by induction in $\ell=a+b$, for example, where $\ell=0,1,\ldots,p+p'-2$. 
Indeed, for $\ell=0$ referring to the left side, 
the equation (\ref{starteq}) reduces to (\ref{idirr}). For higher $\ell$, the right side either vanishes
or involves only terms with lower $\ell$ values (of the form $r+s-2-2i-2i'$)
than the left side where $\ell=r+s-2$, thereby completing the proof of the induction start (\ref{start}).

To establish the induction step in $k'$, we use (\ref{start}) and consider
\be
 \R_{pn,p'}^{a,b}\otimes\R_{1,p'n'}^{0,1}\ =\ \R_{p,p'n}^{a,b}\otimes\R_{1,p'n'}^{0,1}
\label{RRk}
\ee
Since $1\leq p<p'$, the representation $\R_{1,p'n'}^{0,1}$ is well defined for all ${\cal LM}(p,p')$.
Again employing our fusion prescription (\ref{rs}) and the component fusion rules
of Section \ref{onelegged}, we find that the equality (\ref{RRk}) implies that
\bea
 0\!\!&=&\!\! \Big\{\R_{pn,p'(n'+1)}^{a,b}\ominus\Big(\bigoplus_{j=|n-n'-1|+1,\ \!{\rm by}\ \!2}^{n+n'}
    \R_{p,p'j}^{a,b}\Big)\Big\}\nn
 \!\!&\oplus&\!\!\Big(\bigoplus_{i=1}^{\lfloor\frac{b}{2}\rfloor}
  \ \! 2\ \! \Big\{\R_{pn,p'(n'+1)}^{a,b-2i}\ominus\Big(\bigoplus_{j=|n-n'-1|+1,\ \!{\rm by}\ \!2}^{n+n'}
    \R_{p,p'j}^{a,b-2i}\Big)\Big\}\Big)  \nn
 \!\!&\oplus&\!\!\Big(\bigoplus_{i=0}^{\lfloor\frac{b}{2}\rfloor} \big(2-\delta_{i,0}\big)
   \Big\{\R_{pn,p'(n'-1)}^{a,b-2i}\ominus\Big(\bigoplus_{j=|n-n'+1|+1,\ \!{\rm by}\ \!2}^{n+n'-2}
    \R_{p,p'j}^{a,b-2i}\Big)\Big\}\Big)  \nn
 \!\!&\oplus&\!\!\Big(\bigoplus_{i=0}^{\lfloor\frac{p'-b}{2}\rfloor} \big(4-2\delta_{i,0}\big)
   \Big\{\R_{pn,p'n'}^{a,p'-b-2i}\ominus\Big(\bigoplus_{j=|n-n'|+1,\ \!{\rm by}\ \!2}^{n+n'-1}
    \R_{p,p'j}^{a,p'-b-2i}\Big)\Big\}\Big) 
\label{stepeq}
\eea
Here the third and fourth lines vanish by induction assumption. The induction step in $k'$
subsequently follows from (\ref{stepeq}) 
by induction in $b$ in much the same way as the induction start (\ref{start})
followed from (\ref{starteq}) by induction in $\ell=a+b$.

To establish the induction step in $k$, we first assume that $p>1$ in which case the proof
goes as the proof of the induction step in $k'$, this time being based on the equality
\be
 \R_{p,p'n'}^{a,b}\otimes\R_{pn,1}^{1,0}\ =\ \R_{pn',p'}^{a,b}\otimes\R_{pn,1}^{1,0}
\label{RRkk}
\ee
instead of (\ref{RRk}).
For $p=1$, we simply have
\be
 \R_{n,p'n'}^{0,b}\ =\ (n,1)\otimes\R_{1,p'n'}^{0,b}\ =\ (n,1)\otimes\R_{n',p'}^{0,b}
  \ =\ \bigoplus_{j=|n-n'|+1,\ \!{\rm by}\ \!2}^{n+n'-1}\R_{j,p'}^{0,b}
\ee
This concludes the proof of the proposition that the decompositions (\ref{kkexp})
are direct consequences of our fusion prescription.

\subsection{Fundamental Fusion Algebra}

Employing our fusion prescription (\ref{rs}), the fundamental fusion algebra
$\big\langle(2,1),(1,2)\big\rangle_{p,p'}$
now follows straightforwardly from the 
horizontal and vertical fusion algebras $\big\langle(2,1)\big\rangle_{p,p'}$ 
and $\big\langle(1,2)\big\rangle_{p,p'}$ described in Section \ref{onelegged}. 
Let us illustrate this by considering the fusion 
$\R_{pk,p'}^{a,b}\otimes\R_{pk',p'}^{a',b'}$, where $a+a'>p$ and $b+b'\leq p'$,
of two rank-3 indecomposable representations 
\bea
 \R_{pk,p'}^{a,b}\otimes\R_{pk',p'}^{a',b'}\!\!&=&\!\!\Big(\R_{pk,1}^{a,0}\otimes\R_{pk',1}^{a',0}\Big)
   \otimes\Big(\R_{1,p'}^{0,b}\otimes\R_{1,p'}^{0,b'}\Big)
   \ =\ S^{(8)}\oplus S^{(4)}\oplus S^{(2)}\oplus S^{(0)}
\label{33}
\eea   
where
\bea
 S^{(8)}\!\!&=&\!\! 
  \Big(\bigoplus_{j=|k-k'|-2,\ \!{\rm by}\ \!2}^{k+k'+2}\dddkk
    \Big\{\bigoplus_{i=0}^{\lfloor\frac{a+a'-p-1}{2}\rfloor} \nn 
 &&\hspace{0.2cm}  
   \Big(\Big\{\bigoplus_{i'=0}^{\lfloor\frac{|b-b'|-1}{2}\rfloor}\R_{pj,p'}^{a+a'-p-1-2i,|b-b'|-1-2i'}\Big\}
    \oplus\Big\{\bigoplus_{i'=0}^{\lfloor\frac{b+b'-1}{2}\rfloor}\R_{pj,p'}^{a+a'-p-1-2i,b+b'-1-2i'}
      \Big\}\Big)\Big\}\Big)
\eea
\bea
 S^{(4)}\!\!&=&\!\! 
  \Big(\bigoplus_{j=|k-k'|-1,\ \!{\rm by}\ \!2}^{k+k'+1}2\ddkk
    \Big\{\bigoplus_{i=0}^{\lfloor\frac{a+a'-p-1}{2}\rfloor} \nn 
 &&\hspace{0.2cm}  
   \Big(\Big\{\bigoplus_{i'=0}^{\lfloor\frac{p'-|b-b'|-1}{2}\rfloor}\R_{pj,p'}^{a+a'-p-1-2i,p'-|b-b'|-1-2i'}\Big\}
    \oplus\Big\{\bigoplus_{i'=0}^{\lfloor\frac{p'-b-b'-1}{2}\rfloor}\R_{pj,p'}^{a+a'-p-1-2i,p'-b-b'-1-2i'}
      \Big\}\Big)\Big\}\Big)\nn
 \!\!&\oplus&\!\! 
  \Big(\bigoplus_{j=|k-k'|-1,\ \!{\rm by}\ \!2}^{k+k'+1}\ddkk
    \Big\{\bigoplus_{i=0}^{\lfloor\frac{|a-a'|-1}{2}\rfloor} \nn 
 &&\hspace{0.2cm}  
   \Big(\Big\{\bigoplus_{i'=0}^{\lfloor\frac{|b-b'|-1}{2}\rfloor}\R_{pj,p'}^{|a-a'|-1-2i,|b-b'|-1-2i'}\Big\}
    \oplus\Big\{\bigoplus_{i'=0}^{\lfloor\frac{b+b'-1}{2}\rfloor}\R_{pj,p'}^{|a-a'|-1-2i,b+b'-1-2i'}
      \Big\}\Big)\Big\}\Big)\nn
 \!\!&\oplus&\!\! 
  \Big(\bigoplus_{j=|k-k'|-1,\ \!{\rm by}\ \!2}^{k+k'+1}\ddkk
    \Big\{\bigoplus_{i=0}^{\lfloor\frac{2p-a-a'-1}{2}\rfloor} \nn 
 &&\hspace{0.2cm}  
   \Big(\Big\{\bigoplus_{i'=0}^{\lfloor\frac{|b-b'|-1}{2}\rfloor}\R_{pj,p'}^{2p-a-a'-1-2i,|b-b'|-1-2i'}\Big\}
    \oplus\Big\{\bigoplus_{i'=0}^{\lfloor\frac{b+b'-1}{2}\rfloor}\R_{pj,p'}^{2p-a-a'-1-2i,b+b'-1-2i'}
      \Big\}\Big)\Big\}\Big)
\eea
\bea
 S^{(2)}\!\!&=&\!\! 
  \Big(\bigoplus_{j=|k-k'|,\ \!{\rm by}\ \!2}^{k+k'}2\dkk
    \Big\{\bigoplus_{i=0}^{\lfloor\frac{|a-a'|-1}{2}\rfloor} \nn 
 &&\hspace{0.2cm}  
   \Big(\Big\{\bigoplus_{i'=0}^{\lfloor\frac{p'-|b-b'|-1}{2}\rfloor}\R_{pj,p'}^{|a-a'|-1-2i,p'-|b-b'|-1-2i'}\Big\}
    \oplus\Big\{\bigoplus_{i'=0}^{\lfloor\frac{p'-b-b'-1}{2}\rfloor}\R_{pj,p'}^{|a-a'|-1-2i,p'-b-b'-1-2i'}
      \Big\}\Big)\Big\}\Big)\nn
 \!\!&\oplus&\!\! 
  \Big(\bigoplus_{j=|k-k'|,\ \!{\rm by}\ \!2}^{k+k'}2\dkk
    \Big\{\bigoplus_{i=0}^{\lfloor\frac{2p-a-a'-1}{2}\rfloor} \nn 
 &&\hspace{0.2cm}  
   \Big(\Big\{\bigoplus_{i'=0}^{\lfloor\frac{p'-|b-b'|-1}{2}\rfloor}\R_{pj,p'}^{2p-a-a'-1-2i,p'-|b-b'|-1-2i'}\Big\}
    \oplus\Big\{\bigoplus_{i'=0}^{\lfloor\frac{p'-b-b'-1}{2}\rfloor}\R_{pj,p'}^{2p-a-a'-1-2i,p'-b-b'-1-2i'}
      \Big\}\Big)\Big\}\Big)\nn
 \!\!&\oplus&\!\! 
  \Big(\bigoplus_{j=|k-k'|,\ \!{\rm by}\ \!2}^{k+k'}2\dkk
    \Big\{\bigoplus_{i=0}^{p-\max\{a,a'\}-1} \nn 
 &&\hspace{0.2cm}  
   \Big(\Big\{\bigoplus_{i'=0}^{\lfloor\frac{|b-b'|-1}{2}\rfloor}\R_{pj,p'}^{p-|a-a'|-1-2i,|b-b'|-1-2i'}\Big\}
    \oplus\Big\{\bigoplus_{i'=0}^{\lfloor\frac{b+b'-1}{2}\rfloor}\R_{pj,p'}^{p-|a-a'|-1-2i,b+b'-1-2i'}
      \Big\}\Big)\Big\}\Big)
\eea
and
\bea
 S^{(0)}\!\!&=&\!\! 
  \Big(\bigoplus_{j=|k-k'|+1,\ \!{\rm by}\ \!2}^{k+k'-1}4\ \!
    \Big\{\bigoplus_{i=0}^{p-\max\{a,a'\}-1} \nn 
 &&\hspace{0.2cm}  
   \Big(\Big\{\bigoplus_{i'=0}^{\lfloor\frac{p'-|b-b'|-1}{2}\rfloor}\R_{pj,p'}^{p-|a-a'|-1-2i,p'-|b-b'|-1-2i'}\Big\}
    \oplus\Big\{\bigoplus_{i'=0}^{\lfloor\frac{p'-b-b'-1}{2}\rfloor}\R_{pj,p'}^{p-|a-a'|-1-2i,p'-b-b'-1-2i'}
      \Big\}\Big)\Big\}\Big)\nn
\eea
Depending on the relations between the various parameters, the expression (\ref{33}) can of course
be simplified.

The Kac representation $(1,1)$ is the identity of 
the fundamental fusion algebra of ${\cal LM}(p,p')$. To see this, we first argue
that $(1,1)$ is indeed generated by successive fusion of the fundamental Kac representations
$(2,1)$ and $(1,2)$.
For $p'>2$, this follows from the fundamental fusion $(1,2)\otimes(1,2)=(1,1)\oplus(1,3)$,
while for $p'=2$ (in which case $p=1$), 
it follows from the fundamental fusion $(2,1)\otimes(2,1)=(1,1)\oplus(3,1)$.
Letting $X$ denote any representation 
in the algebra, it is easily verified, using the explicit fusion rules for the
fundamental component fusion algebras in Section \ref{onelegged}, that $(1,1)\otimes X=X$,
hence $(1,1)$ is the identity with respect to fusion.
It is also noted that the identity $(1,1)$ is an irreducible representation for $p=1$ but
a reducible yet indecomposable representation of rank 1 for $p>1$.

The fundamental fusion algebra of 
critical percolation ${\cal LM}(2,3)$ was considered in \cite{RP07}
and found to reproduce the many explicit examples of fusion rules 
for the augmented $c_{2,3}$ model appearing in \cite{EF06}.
After discussing an underlying $s\ell(2)$ structure of our fusion rules,
we provide details on the fundamental fusion algebras of the infinite sequence of
logarithmic minimal models ${\cal LM}(1,p')$, the logarithmic Yang-Lee model ${\cal LM}(2,5)$
and the logarithmic Ising model ${\cal LM}(3,4)$. The results for ${\cal LM}(1,p')$
and ${\cal LM}(2,5)$ are subsequently compared with the fusion rules of the
corresponding augmented $c_{p,p'}$ models appearing in the literature \cite{GK96,EF06}.

\subsection{$s\ell(2)$ Structure}

We wish to point out that, at the level of Kac characters, the horizontal, vertical and
fundamental fusion algebras are all compatible with the $s\ell(2)$ structure
\be
 \phi_n\otimes\phi_{n'}\ =\ \bigoplus_{m=|n-n'|+1,\ \!{\rm by}\ \!2}^{n+n'-1}\phi_m
\label{sl2}
\ee
This is straightforward to establish for the horizontal and vertical fusion algebras.
Let us illustrate this by considering the relatively complicated horizontal
fusion $\R_{pk,1}^{a,0}\otimes\R_{pk',1}^{a',0}$ for $a+a'>p$, where (\ref{sl2}) gives
\bea
 \chit[\R_{pk,1}^{a,0}\otimes\R_{pk',1}^{a',0}](q)\!\!&=&\!\!
  \chit\Big[\big((pk-a,1)\oplus(pk+a,1)\big)\otimes\big((pk'-a',1)\oplus(pk'+a',1)\big)\Big](q)\nn
 &=&\!\!\chit\big[(pk-a,1)\otimes(pk'-a',1)\big](q)+\chit\big[(pk-a,1)\otimes(pk'+a',1)\big](q)\nn
 &+&\!\! \chit\big[(pk+a,1)\otimes(pk'-a',1)\big](q)+\chit\big[(pk+a,1)\otimes(pk'+a',1)\big](q)\nn
 \!\!&=&\!\!\sum_{j=|pk-pk'-a+a'|+1,\ \!{\rm by}\ \!2}^{p(k+k')-a-a'-1}\chit_{j,1}(q)
  +\sum_{j=|pk-pk'-a-a'|+1,\ \!{\rm by}\ \!2}^{p(k+k')-a+a'-1}\chit_{j,1}(q)\nn
 \!\!&+&\!\!\sum_{j=|pk-pk'+a+a'|+1,\ \!{\rm by}\ \!2}^{p(k+k')+a-a'-1}\chit_{j,1}(q)
  +\sum_{j=|pk-pk'+a-a'|+1,\ \!{\rm by}\ \!2}^{p(k+k')+a+a'-1}\chit_{j,1}(q)
\label{sl2Rc}
\eea
while (\ref{aap}) yields
\bea
 &&\!\!  \chit[\R_{pk,1}^{a,0}\otimes\R_{pk',1}^{a',0}](q)\nn
 \!\!&=&\!\!  \sum_{j=|k-k'|-1,\ \!{\rm by}\ \!2}^{k+k'+1}\ddkk
    \sum_{i=0}^{\lfloor\frac{a+a'-p-1}{2}\rfloor}
      \big(\chit_{p(j+1)-a-a'+1+2i,1}(q)+\chit_{p(j-1)+a+a'-1-2i,1}(q)\big)\nn
 \!\!&+&\!\!\sum_{j=|k-k'|,\ \!{\rm by}\ \!2}^{k+k'}\dkk
  \sum_{i=0}^{\lfloor\frac{|a-a'|-1}{2}\rfloor}
    \big(\chit_{pj-|a-a'|+1+2i,1}(q)+\chit_{pj+|a-a'|-1-2i,1}(q)\big)\nn
 \!\!&+&\!\!\sum_{j=|k-k'|,\ \!{\rm by}\ \!2}^{k+k'}\dkk
  \sum_{i=0}^{\lfloor\frac{2p-a-a'-1}{2}\rfloor}
    \big(\chit_{p(j-2)+a+a'+1+2i,1}(q)+\chit_{p(j+2)-a-a'-1-2i,1}(q)\big)\nn
 \!\!&+&\!\!
   2\sum_{j=|k-k'|+1,\ \!{\rm by}\ \!2}^{k+k'-1}
    \sum_{i=0}^{p-\max\{a,a'\}-1}\big(\chit_{p(j-1)+|a-a'|+1+2i,1}(q)+\chit_{p(j+1)-|a-a'|-1-2i,1}(q)\big)
\label{sl2R}
\eea
It is straightforward to verify the equality of the two character expressions (\ref{sl2Rc}) and (\ref{sl2R}).
The separation into horizontal and vertical parts then implies that 
the characters of the fundamental fusion algebra exhibit two independent 
$s\ell(2)$ structures as in (\ref{sl2}) --- one in each direction. This is clearly
reminiscent of the fusion algebras of rational (minimal) models where the
$s\ell(2)$ structures are carried by the (characters of the) {\em irreducible} representations.
Here, on the other hand, the $s\ell(2)$ structures are tied to the 
{\em Kac} representations but, due to the higher-rank indecomposable nature of some 
other representations, only at the level of their {\em characters}.

\subsection{Critical Dense Polymers ${\cal LM}(1,2)$ and General ${\cal LM}(1,p')$}

In the case of ${\cal LM}(1,p')$, no indecomposable representation of rank 3 arises when 
combining the horizontal fusion algebra
\be
 \big\langle(2,1)\big\rangle_{1,p'}\ =\  \big\langle(r,1);\ r\in\mathbb{N}\big\rangle_{1,p'}
\ee
with the vertical fusion algebra
\be
 \big\langle(1,2)\big\rangle_{1,p'}\ =\  \big\langle(1,s_0), (1,kp'),\R_{1,kp'}^{0,b};\ s_0,b=1,2,\ldots,p'-1;
   \ k\in\mathbb{N}\big\rangle_{1,p'} 
\ee
The only new representations generated by the merge are the irreducible
Kac representations $(r,s_0)$ with $1<s_0<p'$ as we have
\be
 \big\langle(2,1),(1,2)\big\rangle_{1,p'}\ =\  \big\langle(r,s_0), (1,kp'),\R_{1,kp'}^{0,b};\ 
  s_0,b=1,2,\ldots,p'-1; \ r,k\in\mathbb{N}\big\rangle_{1,p'} 
\label{fund1p}
\ee
This means, in particular, that the fundamental fusion algebra of ${\cal LM}(1,p')$ follows almost trivially
from the fundamental component fusion algebra of order $p'$.
In the special case of critical dense polymers ${\cal LM}(1,2)$, we thus have
\be
  \big\langle(2,1),(1,2)\big\rangle_{1,2}\ =\ \big\langle(r,1), (1,2k),\R_{1,2k}^{0,1};\ 
    r,k\in\mathbb{N}\big\rangle_{1,2} 
\ee
with fusion rules
\bea
 (r,1)\otimes(r',1)\!\!&=&\!\!   \bigoplus_{j=|r-r'|+1,\ \!{\rm by}\ \!2}^{r+r'-1} (j,1)\nn
 (r,1)\otimes(1,2k)\!\!&=&\!\!      \bigoplus_{j=|r-k|+1,\ \!{\rm by}\ \!2}^{r+k-1} (1,2j)\nn
 (r,1)\otimes\R_{1,2k}^{0,1}\!\!&=&\!\!  \bigoplus_{j=|r-k|+1,\ \!{\rm by}\ \!2}^{r+k-1}\R_{1,2j}^{0,1}   \nn
 (1,2k)\otimes(1,2k')\!\!&=&\!\!   \bigoplus_{j=|k-k'|+1,\ \!{\rm by}\ \!2}^{k+k'-1}\R_{1,2j}^{0,1}   \nn
 (1,2k)\otimes\R_{1,2k'}^{0,1}\!\!&=&\!\! \bigoplus_{j=|k-k'|}^{k+k'}\dkk(1,2j)   \nn
 \R_{1,2k}^{0,1}\otimes\R_{1,2k'}^{0,1}\!\!&=&\!\! \bigoplus_{j=|k-k'|}^{k+k'}\dkk\R_{1,2j}^{0,1}  
\label{fund12exp}
\eea

In \cite{GK96}, Gaberdiel and Kausch performed the first systematic analysis of the fusion algebra
of the augmented $c_{1,p'}$ models by application of the Nahm algorithm \cite{Nahm94}.
Based on this, they presented explicit conjectures for the fusion algebras of the augmented
$c_{1,2}$ and $c_{1,3}$ models in addition to a couple of conjectures for fusion rules for $p'>3$. 
To facilitate a comparison of our results with theirs, 
we provide a dictionary for translating the representations generating the 
fundamental fusion algebra of ${\cal LM}(1,p')$ (\ref{fund1p}) into the notation used in \cite{GK96}
\bea
 p'\ \ &\longleftrightarrow&\ \  t\nn
 (r,s_0)\ \ &\longleftrightarrow&\ \  {\cal V}_{r,s_0},\hspace{1.6cm}s_0=1,2,\ldots,p'-1\nn
 (1,kp')=(k,p')\ \ &\longleftrightarrow&\ \ {\cal V}_{k,t}\nn
 \R_{1,kp'}^{0,b}\ \ &\longleftrightarrow&\ \ \R_{k,t-b},\hspace{1.4cm}b=1,2,\ldots,p'-1
\label{dictGK}
\eea
where $r,k\in\mathbb{N}$. It is readily verified that our fusion rules extend and complete the ones
by Gaberdiel and Kausch. In particular, the fusion rules (\ref{fund12exp}) for critical dense polymers
agree exactly with the similar rules in \cite{GK96}.

\subsection{Logarithmic Yang-Lee Model ${\cal LM}(2,5)$}

The fundamental fusion algebra of the logarithmic Yang-Lee model ${\cal LM}(2,5)$
is obtained by combining a fundamental component fusion algebra of order 2 with
a fundamental component fusion algebra of order 5. According to (\ref{A2112}), 
closure of the fusion algebra requires 
\be
 \big\langle(2,1),(1,2)\big\rangle_{2,5}\ =\ \big\langle(1,s_0), (2k,s_0), (1,5k), (2k,5),
   \R_{2k,s_0}^{1,0}, \R_{2k,5}^{1,0}, \R_{1,5k}^{0,b}, \R_{2,5k}^{0,b}  
  \R_{2k,5}^{1,b}\big\rangle_{2,5}
\ee
where $s_0,b=1,2,3,4$ whereas $k\in\mathbb{N}$.

We have already employed our fusion prescription in several examples but let us nevertheless illustrate
its usability again by considering the fusions $\R_{2,3}^{1,0}\otimes\R_{4,5}^{0,3}$ 
and $(1,4)\otimes\R_{8,5}^{1,3}$ in detail.
We find 
\bea
 \R_{2,3}^{1,0}\otimes\R_{4,5}^{0,3}\!\!&=&\!\!
  \big(\R_{2,1}^{1,0}\otimes(4,1)\big)\otimes\big((1,3)\otimes\R_{1,5}^{0,3}\big)
 \ =\ \big((2,1)\oplus2(4,1)\oplus(6,1)\big)\otimes\big(\R_{1,5}^{0,1}\oplus\R_{1,5}^{0,3}
  \oplus(1,10)\big)\nn
 \!\!&=&\!\!\R_{2,5}^{0,1}\oplus\R_{2,5}^{0,3}\oplus2\R_{2,10}^{0,1}\oplus2\R_{2,10}^{0,3}
  \oplus\R_{2,15}^{0,1}\oplus\R_{2,15}^{0,3}\oplus
    2(2,5)\oplus2(4,5)\oplus2(6,5)\oplus(8,5)\nn
\label{YL1}
\eea
and
\bea
 (1,4)\otimes\R_{8,5}^{1,3}\!\!&=&\!\!  \R_{8,1}^{1,0}\otimes\big((1,4)\otimes\R_{1,5}^{0,3}\big)  
  \ =\ \R_{8,1}^{1,0}\otimes\big(2(1,5)\oplus\R_{1,5}^{0,2}\oplus\R_{1,10}^{0,1}\big)\nn
 \!\!&=&\!\!\R_{6,5}^{1,1}\oplus2\R_{8,5}^{1,0}\oplus\R_{8,5}^{1,2}\oplus\R_{10,5}^{1,1}
\label{YL2}
\eea

We now compare our fusion rules for the logarithmic Yang-Lee model ${\cal LM}(2,5)$
with the examples of fusions in the augmented $c_{2,5}$ model considered recently
by Eberle and Flohr \cite{EF06}.
To facilitate such a comparison, we provide a partial dictionary relating
our notation to the one used in \cite{EF06}. In the orders specified, the translation reads
\bea
 \{(2k,s),(1,5k)\}
   &\longleftrightarrow&\{{\cal V}(\D_{2k,s}),{\cal V}(\D_{1,5k})\},
     \hspace{2cm} s=1,2,3,4,5;\ k\in\mathbb{N}\nn
 \{(1,1),(1,2),(1,3),(1,4)\}
   &\longleftrightarrow&\{\R^{(1)}(0)_{4},\R^{(1)}(-1/5)_{3},\R^{(1)}(-1/5)_{2},\R^{(1)}(0)_{1}\}\nn
 \{\R_{2,1}^{1,0},\R_{2,2}^{1,0},\R_{2,3}^{1,0},\R_{2,4}^{1,0},\R_{2,5}^{1,0}\}
  &\longleftrightarrow&\{\R^{(2)}(0,4)_{13},\R^{(2)}(-1/5,14/5)_{11},\R^{(2)}(-1/5,9/5)_{9},\nn
     &&\ \ \ \R^{(2)}(0,1)_{7},\R^{(2)}(2/5,2/5)\}\nn
 \{\R_{1,5}^{0,1},\R_{1,5}^{0,2},\R_{1,5}^{0,3},\R_{1,5}^{0,4}\}
  &\longleftrightarrow&\{\R^{(2)}(0,1)_{13},\R^{(2)}(-1/5,9/5)_{11},
    \R^{(2)}(-1/5,14/5)_{9},\R^{(2)}(0,4)_{7}\}\nn
 \{\R_{2,5}^{0,1},\R_{2,5}^{0,2},\R_{2,5}^{0,3},\R_{2,5}^{0,4}\}
  &\longleftrightarrow&\{\R^{(2)}(-1/8,-1/8),\R^{(2)}(7/40,7/40),\nn
    &&\ \ \ \R^{(2)}(27/40,27/40),\R^{(2)}(11/8,11/8)\}\nn
 \{\R_{2,5}^{1,1},\R_{2,5}^{1,2},\R_{2,5}^{1,3},\R_{2,5}^{1,4}\}
  &\longleftrightarrow&\{\R^{(3)}(0,0,1,1),\R^{(3)}(-1/5,-1/5,9/5,9/5),\nn
    &&\ \ \ \R^{(3)}(-1/5,-1/5,14/5,14/5),\R^{(3)}(0,0,4,4)\}
\label{dictEF}
\eea
We find that our fusion rules reproduce the many explicit examples considered in \cite{EF06}.
As our rules are general, the fusions of the four indecomposable representations of rank 3
appearing in the dictionary (\ref{dictEF}) are easily worked out to be
\bea
 \R_{2,5}^{1,1}\otimes\R_{2,5}^{1,1}\!\!&=&\!\!  
  8\R_{2,5}^{1,0}\oplus\R_{2,5}^{1,1}\oplus8\R_{2,5}^{1,2}\oplus4\R_{2,5}^{1,4}\oplus4\R_{4,5}^{1,0}
  \oplus2\R_{4,5}^{1,1}\oplus4\R_{4,5}^{1,2}\oplus2\R_{4,5}^{1,4}\oplus\R_{6,5}^{1,1}  \nn
 \R_{2,5}^{1,1}\otimes\R_{2,5}^{1,2}\!\!&=&\!\!  
  2\R_{2,5}^{1,0}\oplus8\R_{2,5}^{1,1}\oplus\R_{2,5}^{1,2}\oplus4\R_{2,5}^{1,3}\oplus4\R_{4,5}^{1,0}
  \oplus4\R_{4,5}^{1,1}\oplus2\R_{4,5}^{1,2}\oplus2\R_{4,5}^{1,3}\oplus2\R_{6,5}^{1,0}
  \oplus\R_{6,5}^{1,2}  \nn
 \R_{2,5}^{1,1}\otimes\R_{2,5}^{1,3}\!\!&=&\!\!   
  8\R_{2,5}^{1,0}\oplus2\R_{2,5}^{1,1}\oplus4\R_{2,5}^{1,2}\oplus\R_{2,5}^{1,3}\oplus4\R_{4,5}^{1,0}
  \oplus4\R_{4,5}^{1,1}\oplus2\R_{4,5}^{1,2}\oplus2\R_{4,5}^{1,3}\oplus2\R_{6,5}^{1,1}
  \oplus\R_{6,5}^{1,3}  \nn
 \R_{2,5}^{1,1}\otimes\R_{2,5}^{1,4}\!\!&=&\!\!   
  2\R_{2,5}^{1,0}\oplus4\R_{2,5}^{1,1}\oplus2\R_{2,5}^{1,2}\oplus\R_{2,5}^{1,4}\oplus4\R_{4,5}^{1,0}
  \oplus2\R_{4,5}^{1,1}\oplus4\R_{4,5}^{1,2}\oplus2\R_{4,5}^{1,4}\nn
 \!\!&\oplus&\!\!2\R_{6,5}^{1,0}
  \oplus2\R_{6,5}^{1,2}\oplus\R_{6,5}^{1,4}  \nn
 \R_{2,5}^{1,2}\otimes\R_{2,5}^{1,2}\!\!&=&\!\!    
  8\R_{2,5}^{1,0}\oplus\R_{2,5}^{1,1}\oplus4\R_{2,5}^{1,2}\oplus\R_{2,5}^{1,3}\oplus4\R_{2,5}^{1,4}
  \oplus4\R_{4,5}^{1,0}
  \oplus2\R_{4,5}^{1,1}\oplus2\R_{4,5}^{1,2}\oplus2\R_{4,5}^{1,3}\oplus2\R_{4,5}^{1,4}\nn
 \!\!&\oplus&\!\!\R_{6,5}^{1,1}\oplus\R_{6,5}^{1,3}  \nn
 \R_{2,5}^{1,2}\otimes\R_{2,5}^{1,3}\!\!&=&\!\!   
  2\R_{2,5}^{1,0}\oplus4\R_{2,5}^{1,1}\oplus\R_{2,5}^{1,2}\oplus4\R_{2,5}^{1,3}\oplus\R_{2,5}^{1,4}
  \oplus4\R_{4,5}^{1,0}
  \oplus2\R_{4,5}^{1,1}\oplus2\R_{4,5}^{1,2}\oplus2\R_{4,5}^{1,3}\oplus2\R_{4,5}^{1,4}\nn
 \!\!&\oplus&\!\!2\R_{6,5}^{1,0}\oplus\R_{6,5}^{1,2}\oplus\R_{6,5}^{1,4}  \nn
 \R_{2,5}^{1,2}\otimes\R_{2,5}^{1,4}\!\!&=&\!\!   
  2\R_{2,5}^{1,0}\oplus2\R_{2,5}^{1,1}\oplus4\R_{2,5}^{1,2}\oplus\R_{2,5}^{1,3}
  \oplus2\R_{4,5}^{1,0}
  \oplus4\R_{4,5}^{1,1}\oplus2\R_{4,5}^{1,2}\oplus2\R_{4,5}^{1,3}\nn
 \!\!&\oplus&\!\!2\R_{6,5}^{1,0}\oplus2\R_{6,5}^{1,1}\oplus\R_{6,5}^{1,3}\oplus\R_{8,5}^{1,0}  \nn
 \R_{2,5}^{1,3}\otimes\R_{2,5}^{1,3}\!\!&=&\!\!    
  2\R_{2,5}^{1,0}\oplus\R_{2,5}^{1,1}\oplus4\R_{2,5}^{1,2}\oplus\R_{2,5}^{1,3}\oplus4\R_{2,5}^{1,4}
  \oplus2\R_{4,5}^{1,0}
  \oplus2\R_{4,5}^{1,1}\oplus2\R_{4,5}^{1,2}\oplus2\R_{4,5}^{1,3}\oplus2\R_{4,5}^{1,4}\nn
 \!\!&\oplus&\!\!2\R_{6,5}^{1,0}\oplus\R_{6,5}^{1,1}\oplus\R_{6,5}^{1,3}\oplus\R_{8,5}^{1,0}  \nn
 \R_{2,5}^{1,3}\otimes\R_{2,5}^{1,4}\!\!&=&\!\!   
  2\R_{2,5}^{1,0}\oplus2\R_{2,5}^{1,1}\oplus\R_{2,5}^{1,2}\oplus4\R_{2,5}^{1,3}
  \oplus4\R_{4,5}^{1,0}
  \oplus2\R_{4,5}^{1,1}\oplus2\R_{4,5}^{1,2}\oplus2\R_{4,5}^{1,3}\nn
 \!\!&\oplus&\!\!2\R_{6,5}^{1,0}\oplus2\R_{6,5}^{1,1}\oplus\R_{6,5}^{1,2}\oplus\R_{8,5}^{1,1}  \nn
 \R_{2,5}^{1,4}\otimes\R_{2,5}^{1,4}\!\!&=&\!\!    
  2\R_{2,5}^{1,0}\oplus\R_{2,5}^{1,1}\oplus2\R_{2,5}^{1,2}\oplus4\R_{2,5}^{1,4}
  \oplus2\R_{4,5}^{1,0}
  \oplus2\R_{4,5}^{1,1}\oplus\R_{4,5}^{1,2}\oplus2\R_{4,5}^{1,4}\nn
 \!\!&\oplus&\!\!2\R_{6,5}^{1,0}\oplus\R_{6,5}^{1,1}\oplus2\R_{6,5}^{1,2}\oplus\R_{8,5}^{1,0}
  \oplus\R_{8,5}^{1,2}  
\eea
As (\ref{YL1}) and (\ref{YL2}), these explicit fusions were not considered in \cite{EF06}.

\subsection{Logarithmic Ising Model ${\cal LM}(3,4)$ and Beyond}

The fundamental fusion algebra of the logarithmic Ising model ${\cal LM}(3,4)$
is obtained by combining a fundamental component fusion algebra of order 3 with
a fundamental component fusion algebra of order 4. According to (\ref{A2112}), 
closure of the fusion algebra requires 
\be
 \big\langle(2,1),(1,2)\big\rangle_{3,4}\ =\ \big\langle(r_0,s_0), (3k,s_0), (r_0,4k), (3k,4),
   \R_{3k,s_0}^{a,0}, \R_{3k,4}^{a,0}, \R_{r_0,4k}^{0,b}, \R_{3,4k}^{0,b}  
  \R_{3k,4}^{a,b}\big\rangle_{3,4}
\ee
where $r_0,a=1,2$ and $s_0,b=1,2,3$ whereas $k\in\mathbb{N}$.
Writing out the complete set of fusion rules is cumbersome and does not provide
any new insight over the general fusion prescription given above.
Here we therefore only present simplifications of the inequivalent fusions
of the type (\ref{33}) where we find
\bea
 \R_{3k,4}^{2,1}\otimes\R_{3k',4}^{2,1}\!\!&=&\!\!  
    \Big(\bigoplus_{j=|k-k'|-2,\ \!{\rm by}\ \!2}^{k+k'+2}\dddkk\R_{3j,4}^{0,1}\Big)
   \oplus  \Big(\bigoplus_{j=|k-k'|-1,\ \!{\rm by}\ \!2}^{k+k'+1}\ddkk
   \big(4\R_{3j,4}^{0,1}\oplus2\R_{3j,4}^{0,3}\oplus\R_{3j,4}^{1,1}\big)\Big)   \nn
 \!\!&\oplus&\!\!  \Big(\bigoplus_{j=|k-k'|,\ \!{\rm by}\ \!2}^{k+k'}2\dkk
   \big(2\R_{3j,4}^{1,1}\oplus\R_{3j,4}^{1,3}\oplus\R_{3j,4}^{2,1}\big)\Big)\nn
 \!\!&\oplus&\!\!\Big(\bigoplus_{j=|k-k'|+1,\ \!{\rm by}\ \!2}^{k+k'-1}
  4\big(2\R_{3j,4}^{2,1}\oplus\R_{3j,4}^{2,3}\big)\Big)
\eea
\bea
 \R_{3k,4}^{2,1}\otimes\R_{3k',4}^{2,2}\!\!&=&\!\! 
  \Big(\bigoplus_{j=|k-k'|-2}^{k+k'+2}\dddkk
   \big(2\R_{3j,4}^{0,0}\oplus\R_{3j,4}^{0,2}\big)\Big)
  \oplus\Big(\bigoplus_{j=|k-k'|-1}^{k+k'+1}\ddkk
   \big(2\R_{3j,4}^{1,0}\oplus\R_{3j,4}^{1,2}\big)\Big)   \nn
 \!\!&\oplus&\!\!\Big(\bigoplus_{j=|k-k'|}^{k+k'}2\dkk
   \big(2\R_{3j,4}^{2,0}\oplus\R_{3j,4}^{2,2}\big)\Big)
\eea
\bea
 \R_{3k,4}^{2,1}\otimes\R_{3k',4}^{2,3}\!\!&=&\!\! 
   \Big(\bigoplus_{j=|k-k'|-2,\ \!{\rm by}\ \!2}^{k+k'+2}\dddkk
   \big(2\R_{3j,4}^{0,1}\oplus\R_{3j,4}^{0,3}\big)\Big)\nn
 \!\!&\oplus&\!\!  \Big(\bigoplus_{j=|k-k'|-1,\ \!{\rm by}\ \!2}^{k+k'+1}\ddkk
   \big(2\R_{3j,4}^{0,1}\oplus2\R_{3j,4}^{1,1}\oplus\R_{3j,4}^{1,3}\big)\Big)   \nn
 \!\!&\oplus&\!\!  \Big(\bigoplus_{j=|k-k'|,\ \!{\rm by}\ \!2}^{k+k'}2\dkk
   \big(\R_{3j,4}^{1,1}\oplus2\R_{3j,4}^{2,1}\oplus\R_{3j,4}^{2,3}\big)\Big) 
 \oplus\Big(\bigoplus_{j=|k-k'|+1,\ \!{\rm by}\ \!2}^{k+k'-1}4\R_{3j,4}^{2,1}\Big)
\eea
and
\bea
 \R_{3k,4}^{2,2}\otimes\R_{3k',4}^{2,2}\!\!&=&\!\!  
  \Big(\bigoplus_{j=|k-k'|-2}^{k+k'+2}\dddkk
   \big(2\R_{3j,4}^{0,1}\oplus\R_{3j,4}^{0,3}\big)\Big)
  \oplus\Big(\bigoplus_{j=|k-k'|-1}^{k+k'+1}\ddkk
   \big(2\R_{3j,4}^{1,1}\oplus\R_{3j,4}^{1,3}\big)\Big)   \nn
 \!\!&\oplus&\!\!\Big(\bigoplus_{j=|k-k'|}^{k+k'}2\dkk
   \big(2\R_{3j,4}^{2,1}\oplus\R_{3j,4}^{2,3}\big)\Big)
\eea
It is noted that only some of the direct sums appearing in these expressions are in steps of 2.

The main new feature associated to the fundamental fusion algebras of the logarithmic minimal
models ${\cal LM}(p,p')$ for $p>3$ compared to the properties already encountered in the various
models above with $p=1,2,3$ is the appearance of indecomposable representations of rank 3 
as the result of fusion of two reducible Kac representations. 
This occurs in the fusion of $(r_0,s_0)$ and $(r'_0,s'_0)$ if and only if
$r_0+r'_0>p+1$ and $s_0+s'_0>p'+1$ (which indeed requires $p>3$ since $r_0,r'_0<p$).
In this case, we have
\bea
 (r_0,s_0)\otimes(r'_0,s'_0)\!\!&=&\!\! 
  \Big(\bigoplus_{j=|r_0-r'_0|+1,\ \!{\rm by}\ \!2}^{2p-r_0-r'_0-1}\ \ 
    \bigoplus_{j'=|s_0-s'_0|+1,\ \!{\rm by}\ \!2}^{2p'-s_0-s'_0-1}(j,j')\Big)\nn
  \!\!&\oplus&\!\!   
   \Big(\bigoplus_{j=|r_0-r'_0|+1,\ \!{\rm by}\ \!2}^{2p-r_0-r'_0-1}\ \ 
    \bigoplus_{i'=1}^{\lfloor\frac{s_0+s'_0-p'-1}{2}\rfloor}
    \R_{j,p'}^{0,s_0+s'_0-p'-1-2i'}\Big)\nn
  \!\!&\oplus&\!\!  
    \Big(\bigoplus_{i=1}^{\lfloor\frac{r_0+r'_0-p-1}{2}\rfloor}\
      \bigoplus_{j'=|s_0-s'_0|+1,\ \!{\rm by}\ \!2}^{2p'-s_0-s'_0-1}
       \R_{p,j'}^{r_0+r'_0-p-1-2i,0}\Big)  \nn
  \!\!&\oplus&\!\!
    \Big(\bigoplus_{i=1}^{\lfloor\frac{r_0+r'_0-p-1}{2}\rfloor}\
    \bigoplus_{i'=1}^{\lfloor\frac{s_0+s'_0-p'-1}{2}\rfloor}
    \R_{p,p'}^{r_0+r'_0-p-1-2i,s_0+s'_0-p'-1-2i'}\Big)
\eea
where the last line corresponds to a non-vanishing direct sum of indecomposable representations
of rank 3.

It would of course be interesting to compare our fusion rules for the logarithmic Ising 
model ${\cal LM}(3,4)$, in particular, with the fusion rules obtained by application
of the Nahm algorithm \cite{Nahm94} to the augmented $c_{3,4}$ model along the lines
of \cite{GK96,EF06}. If affirmative, such a comparison would provide further evidence to
support the supposition that the augmented $c_{p,p'}$ model and the logarithmic minimal model
${\cal LM}(p,p')$ are equivalent.

\section{Conclusion}

We have presented explicit conjectures for the chiral fundamental
fusion algebras of the logarithmic minimal models 
${\cal LM}(p,p')$. The fusion rules are quasi-rational~\cite{Nahm94} in the sense that the fusion of a finite number of 
representations decomposes into a finite direct sum of representations. 
The fusion rules are also commutative, associative and exhibit an $s\ell(2)$ structure. 
Detailed comparisons of our fusion rules have shown agreement
with the previous results of Gaberdiel and Kausch for $p=1$ and with Eberle and Flohr for $(p,p')=(2,3),
(2,5)$ corresponding to critical percolation 
(where the explicit comparison was carried out in \cite{RP07}) 
and the logarithmic Yang-Lee model, respectively. In the latter cases, we confirm 
that indecomposable representations of rank 3 arise as the result of certain lower-rank fusions. 
We also find that closure of a fundamental fusion algebra is achieved
without the introduction of indecomposable representations of rank higher than 3.
In general, the identity of the fundamental fusion algebra of ${\cal LM}(p,p')$
is a reducible yet indecomposable Kac representation of rank 1.
The conjectured fusion rules are supported, within our lattice approach introduced in \cite{PRZ}, 
by extensive numerical studies of  
the associated integrable lattice models. Details of our lattice findings and numerical results
will be presented elsewhere. Importantly, the agreement of our results with previous results from the algebraic approach lends 
considerable support for the supposition that the logarithmic CFTs $c_{p,p'}$ and ${\cal LM}(p,p')$ should be identitifed. Finally, we intend to consider the full fusion algebra
$\big\langle (2,1),(p+1,1),(1,2),(1,p'+1)\big\rangle_{p,p'}$ of ${\cal LM}(p,p')$ elsewhere. 
It contains the fundamental fusion algebra $\big\langle (2,1),(1,2)\big\rangle_{p,p'}$
as a subalgebra and is `full' in the sense that it involves all Kac representations $(r,s)$ where 
$r,s\in\mathbb{N}$.
\vskip.5cm
\section*{Acknowledgments}
\vskip.1cm
\noindent
This work is supported by the Australian Research Council.

\appendix

\section{Component Fusion Algebras of Low Order}
\label{appEX}

For given order $p=1,2,3,4,5$, 
focus here is on the fusion algebra generated by $\{\R_{pn}^a;\ a=0,1,\ldots,p-1\}$, that is,
on the subalgebra of the fundamental component fusion algebra
generated by all representations but the reducible yet indecomposable representations of rank 1.
\\[.2cm]
\noindent {\bf Order $p=1$}
\be
 \R_{n}^0\otimes\R_{n'}^0\ =\ \bigoplus_{j=|n-n'|+1,\ \!{\rm by}\ \!2}^{n+n'-1}\R_{j}^0
\label{p1}
\ee
\\[.2cm]
\noindent {\bf Order $p=2$}
\bea
 \R_{2n}^0\otimes\R_{2n'}^0\!\!&=&\!\!\bigoplus_{j=|n-n'|+1,\ \!{\rm by}\ \!2}^{n+n'-1}
  \R_{2j}^1\nn
 \R_{2n}^0\otimes \R_{2n'}^{1}\!\!&=&\!\!\bigoplus_{j=|n-n'|}^{n+n'}
  \dnn\R_{2j}^0
\eea
\bea
 \R_{2n}^{1}\otimes \R_{2n'}^{1}\!\!&=&\!\!\bigoplus_{j=|n-n'|}^{n+n'}
  \dnn\R_{2j}^{1}
\label{p2}
\eea
\\[.2cm]
\noindent {\bf Order $p=3$}
\bea
 \R_{3n}^0\otimes\R_{3j}^0\!\!&=&\!\!
  \bigoplus_{j=|n-n'|+1,\ \!{\rm by}\ \!2}^{n+n'-1}\big(\R_{3j}^{2}\oplus\R_{3j}^0\big)\nn
 \R_{3n}^0\otimes \R_{3n'}^{1}\!\!&=&\!\!
  \Big(\bigoplus_{j=|n-n'|,\ \!{\rm by}\ \!2}^{n+n'}\dnn \R_{3j}^0\Big)
  \oplus\Big(\bigoplus_{j=|n-n'|+1,\ \!{\rm by}\ \!2}^{n+n'-1}2\R_{3j}^{1}\Big)\nn
 \R_{3n}^0\otimes \R_{3n'}^{2}\!\!&=&\!\!  
  \Big(\bigoplus_{j=|n-n'|,\ \!{\rm by}\ \!2}^{n+n'}\dnn\R_{3j}^{1}\Big)
  \oplus\Big(\bigoplus_{j=|n-n'|+1,\ \!{\rm by}\ \!2}^{n+n'-1}2\R_{3j}^0\Big)
\eea
\bea
 \R_{3n}^{1}\otimes \R_{3n'}^{1}\!\!&=&\!\!\ 
  \Big(\bigoplus_{j=|n-n'|,\ \!{\rm by}\ \!2}^{n+n'}\dnn\R_{3j}^{1}\Big)
  \oplus\Big(\bigoplus_{j=|n-n'|+1,\ \!{\rm by}\ \!2}^{n+n'-1}\big(2\R_{3j}^{2}\oplus4\R_{3j}^0\big)\Big)\nn
 \R_{3n}^{1}\otimes \R_{3n'}^{2}\!\!&=&\!\!
  \Big(\bigoplus_{j=|n-n'|,\ \!{\rm by}\ \!2}^{n+n'}\dnn\big(\R_{3j}^{2}
   \oplus2\R_{3j}^0\big)\Big)
  \oplus\Big(\bigoplus_{j=|n-n'|+1,\ \!{\rm by}\ \!2}^{n+n'-1}2\R_{3j}^{1}\Big)
\eea
\bea
 \R_{3n}^{2}\otimes R_{3n'}^{2}\!\!&=&\!\! 
 \Big(\bigoplus_{j=|n-n'|-1,\ \!{\rm by}\ \!2}^{n+n'+1}\ddnn\R_{3j}^0\Big)
 \oplus\Big(\bigoplus_{j=|n-n'|,\ \!{\rm by}\ \!2}^{n+n'}\dnn\R_{3j}^{1}\Big)
  \oplus\Big(\bigoplus_{j=|n-n'|+1,\ \!{\rm by}\ \!2}^{n+n'-1}2\R_{3j}^{2}\Big)\nn
\label{p3}
\eea
\\[.2cm]
\noindent {\bf Order $p=4$}
\bea
  \R_{4n}^0\otimes\R_{4n'}^0\!\!&=&\!\!\bigoplus_{j=|n-n'|+1,\ \!{\rm by}\ \!2}^{n+n'-1}
   \big(\R_{4j}^1\oplus\R_{4j}^3\big)\nn
 \R_{4n}^0\otimes\R_{4n'}^1\!\!&=&\!\!
  \Big(\bigoplus_{j=|n-n'|}^{n+n'}\delta_{j,\{n,n'\}}^{(2)}\R_{4j}^0\Big)
  \oplus\Big(\bigoplus_{j=|n-n'|+1,\ \!{\rm by}\ \!2}^{n+n'-1}2\R_{4j}^2\Big)\nn
 \R_{4n}^0\otimes\R_{4n'}^2\!\!&=&\!\!\bigoplus_{j=|n-n'|}^{n+n'}\delta_{j,\{n,n'\}}^{(2)}\R_{4j}^1\nn
 \R_{4n}^0\otimes\R_{4n'}^3\!\!&=&\!\!
  \Big(\bigoplus_{j=|n-n'|}^{n+n'}\delta_{j,\{n,n'\}}^{(2)}\R_{4j}^0\Big)
  \oplus\Big(\bigoplus_{j=|n-n'|,\ \!{\rm by}\ \!2}^{n+n'}\delta_{j,\{n,n'\}}^{(2)}\R_{4j}^2\Big)
\eea
\bea
 \R_{4n}^1\otimes\R_{4n'}^1\!\!&=&\!\!
  \Big(\bigoplus_{j=|n-n'|}^{n+n'}\delta_{j,\{n,n'\}}^{(2)}\R_{4j}^1\Big)
  \oplus\Big(\bigoplus_{j=|n-n'|+1,\ \!{\rm by}\ \!2}^{n+n'-1}\big(2\R_{4j}^1\oplus2\R_{4j}^3\big)\Big)\nn
 \R_{4n}^1\otimes\R_{4n'}^2\!\!&=&\!\!\bigoplus_{j=|n-n'|}^{n+n'}\delta_{j,\{n,n'\}}^{(2)}\big(2\R_{4j}^0
    \oplus\R_{4j}^2\big)         \nn
 \R_{4n}^1\otimes\R_{4n'}^3\!\!&=&\!\!
  \Big(\bigoplus_{j=|n-n'|}^{n+n'}\delta_{j,\{n,n'\}}^{(2)}\R_{4j}^1\Big)
    \oplus\Big(\bigoplus_{j=|n-n'|,\ \!{\rm by}\ \!2}^{n+n'}\delta_{j,\{n,n'\}}^{(2)}\big(\R_{4j}^1
       \oplus\R_{4j}^3\big)\Big)         
\eea
\bea
 \R_{4n}^2\otimes\R_{4n'}^2\!\!&=&\!\!  \bigoplus_{j=|n-n'|}^{n+n'}
   \delta_{j,\{n,n'\}}^{(2)}\big(\R_{4j}^1\oplus\R_{4j}^3\big)       \nn
 \R_{4n}^2\otimes\R_{4n'}^3\!\!&=&\!\!
  \Big(\bigoplus_{j=|n-n'|-1}^{n+n'+1}\delta_{j,\{n,n'\}}^{(4)}\R_{4j}^0\Big)
   \oplus\Big(\bigoplus_{j=|n-n'|}^{n+n'}\dnn\R_{4j}^2\Big)        
\eea
\bea
 \R_{4n}^3\otimes\R_{4n'}^3\!\!&=&\!\! 
  \Big(\bigoplus_{j=|n-n'|-1,\ \!{\rm by}\ \!2}^{n+n'+1}\ddnn\R_{4j}^1\Big)
  \oplus\Big(\bigoplus_{j=|n-n'|,\ \!{\rm by}\ \!2}^{n+n'}\dnn\R_{4j}^1\Big)
  \oplus\Big(\bigoplus_{j=|n-n'|+1,\ \!{\rm by}\ \!2}^{n+n'-1}2\R_{4j}^3\Big)\nn
\eea
\\[.2cm]
\noindent {\bf Order $p=5$}
\bea
 \R_{5n}^0\otimes\R_{5n'}^0\!\!&=&\!\!\bigoplus_{j=|n-n'|+1,\ \!{\rm by}\ \!2}^{n+n'-1}\big(\R_{5j}^0\oplus
  \R_{5j}^2\oplus\R_{5j}^4\big)\nn
 \R_{5n}^0\otimes\R_{5n'}^1\!\!&=&\!\!
  \Big(\bigoplus_{j=|n-n'|,\ \!{\rm by}\ \!2}^{n+n'}\dnn\R_{5j}^0\Big)
  \oplus\Big( \bigoplus_{j=|n-n'|+1,\ \!{\rm by}\ \!2}^{n+n'-1}\big(2\R_{5j}^1\oplus2\R_{5j}^3\big)\Big)\nn
 \R_{5n}^0\otimes\R_{5n'}^2\!\!&=&\!\!
  \Big(\bigoplus_{j=|n-n'|,\ \!{\rm by}\ \!2}^{n+n'}\dnn\R_{5j}^1\Big)
  \oplus \Big(\bigoplus_{j=|n-n'|+1,\ \!{\rm by}\ \!2}^{n+n'-1}\big(2\R_{5j}^0\oplus2\R_{5j}^2\big)\Big)\nn
 \R_{5n}^0\otimes\R_{5n'}^3\!\!&=&\!\!
  \Big(\bigoplus_{j=|n-n'|,\ \!{\rm by}\ \!2}^{n+n'}\dnn\big(\R_{5j}^0\oplus\R_{5j}^2\big)\Big)
  \oplus\Big(\bigoplus_{j=|n-n'|+1,\ \!{\rm by}\ \!2}^{n+n'-1}2\R_{5j}^1\Big)\nn
 \R_{5n}^0\otimes\R_{5n'}^4\!\!&=&\!\!
  \Big(\bigoplus_{j=|n-n'|,\ \!{\rm by}\ \!2}^{n+n'}\dnn\big(\R_{5j}^1\oplus\R_{5j}^3\big)\Big)
  \oplus\Big(\bigoplus_{j=|n-n'|+1,\ \!{\rm by}\ \!2}^{n+n'-1}2\R_{5j}^0\Big)
\eea
\bea
 \R_{5n}^1\otimes\R_{5n'}^1\!\!&=&\!\!  
   \Big(\bigoplus_{j=|n-n'|,\ \!{\rm by}\ \!2}^{n+n'}\dnn\R_{5j}^1\Big)
   \oplus \Big(\bigoplus_{j=|n-n'|+1,\ \!{\rm by}\ \!2}^{n+n'-1}
     \big(4\R_{5j}^0\oplus4\R_{5j}^2\oplus2\R_{5j}^4\big)\Big)  \nn
 \R_{5n}^1\otimes\R_{5n'}^2\!\!&=&\!\!    
   \Big(\bigoplus_{j=|n-n'|,\ \!{\rm by}\ \!2}^{n+n'}\dnn\big(2\R_{5j}^0\oplus\R_{5j}^2\big)\Big)
   \oplus \Big(\bigoplus_{j=|n-n'|+1,\ \!{\rm by}\ \!2}^{n+n'-1}
     \big(4\R_{5j}^1\oplus2\R_{5j}^3\big)\Big)  \nn
 \R_{5n}^1\otimes\R_{5n'}^3\!\!&=&\!\!  
   \Big(\bigoplus_{j=|n-n'|,\ \!{\rm by}\ \!2}^{n+n'}\dnn\big(2\R_{5j}^1\oplus\R_{5j}^3\big)\Big)
   \oplus \Big(\bigoplus_{j=|n-n'|+1,\ \!{\rm by}\ \!2}^{n+n'-1}
     \big(4\R_{5j}^0\oplus2\R_{5j}^2\big)\Big)  \nn
 \R_{5n}^1\otimes\R_{5n'}^4\!\!&=&\!\!  
   \Big(\bigoplus_{j=|n-n'|,\ \!{\rm by}\ \!2}^{n+n'}\dnn\big(2\R_{5j}^0\oplus2\R_{5j}^2
     \oplus\R_{5j}^4\big)\Big)
   \oplus \Big(\bigoplus_{j=|n-n'|+1,\ \!{\rm by}\ \!2}^{n+n'-1} 2\R_{5j}^1\Big)
\eea
\bea
 \R_{5n}^2\otimes\R_{5n'}^2\!\!&=&\!\!   
   \Big(\bigoplus_{j=|n-n'|,\ \!{\rm by}\ \!2}^{n+n'}\dnn\big(\R_{5j}^1\oplus\R_{5j}^3\big)\Big)
   \oplus \Big(\bigoplus_{j=|n-n'|+1,\ \!{\rm by}\ \!2}^{n+n'-1}
    \big(4\R_{5j}^0\oplus2\R_{5j}^2\oplus2\R_{5j}^4\big)\Big)  \nn
 \R_{5n}^2\otimes\R_{5n'}^3\!\!&=&\!\!   
   \Big(\bigoplus_{j=|n-n'|,\ \!{\rm by}\ \!2}^{n+n'}\dnn\big(2\R_{5j}^0\oplus\R_{5j}^2
     \oplus\R_{5j}^4\big)\Big)
   \oplus \Big(\bigoplus_{j=|n-n'|+1,\ \!{\rm by}\ \!2}^{n+n'-1}
    \big(2\R_{5j}^1\oplus2\R_{5j}^3\big)\Big)  \nn
 \R_{5n}^2\otimes\R_{5n'}^4\!\!&=&\!\!   
   \Big(\bigoplus_{j=|n-n'|-1,\ \!{\rm by}\ \!2}^{n+n'+1}\ddnn\R_{5j}^0\Big)
   \oplus \Big(\bigoplus_{j=|n-n'|,\ \!{\rm by}\ \!2}^{n+n'}\dnn\big(2\R_{5j}^1\oplus\R_{5j}^3\big)\Big)\nn
 \!\!&\oplus&\!\!\Big(\bigoplus_{j=|n-n'|+1,\ \!{\rm by}\ \!2}^{n+n'-1} 2\R_{5j}^2\Big)
\eea
\bea
 \R_{5n}^3\otimes\R_{5n'}^3\!\!&=&\!\!  
   \Big(\bigoplus_{j=|n-n'|-1,\ \!{\rm by}\ \!2}^{n+n'+1}\ddnn\R_{5j}^0\Big)
   \oplus \Big(\bigoplus_{j=|n-n'|,\ \!{\rm by}\ \!2}^{n+n'}\dnn\big(\R_{5j}^1\oplus\R_{5j}^3\big)\Big)\nn
 \!\!&\oplus&\!\!
   \Big(\bigoplus_{j=|n-n'|+1,\ \!{\rm by}\ \!2}^{n+n'-1}\big(2\R_{5j}^2\oplus2\R_{5j}^4\big)\Big) \nn
 \R_{5n}^3\otimes\R_{5n'}^4\!\!&=&\!\!  
   \Big(\bigoplus_{j=|n-n'|-1,\ \!{\rm by}\ \!2}^{n+n'+1}\ddnn\R_{5j}^1\Big)
   \oplus \Big(\bigoplus_{j=|n-n'|,\ \!{\rm by}\ \!2}^{n+n'}\dnn\big(2\R_{5j}^0\oplus\R_{5j}^2\big)\Big)\nn
 \!\!&\oplus&\!\!\Big(\bigoplus_{j=|n-n'|+1,\ \!{\rm by}\ \!2}^{n+n'-1}2\R_{5j}^3\Big)
\eea
\bea
 \R_{5n}^4\otimes\R_{5n'}^4\!\!&=&\!\!    
   \Big(\bigoplus_{j=|n-n'|-1,\ \!{\rm by}\ \!2}^{n+n'+1}\ddnn\big(\R_{5j}^0\oplus\R_{5j}^2\big)\Big)
   \oplus \Big(\bigoplus_{j=|n-n'|,\ \!{\rm by}\ \!2}^{n+n'}\dnn\R_{5j}^1\Big)\nn
\!\!&\oplus&\!\!\Big(\bigoplus_{j=|n-n'|+1,\ \!{\rm by}\ \!2}^{n+n'-1}2\R_{5j}^4\Big) 
\eea


\end{document}